\documentclass[unnumsec,webpdf,contemporary,large]{article}
\usepackage[english]{babel}

\usepackage[letterpaper,top=2cm,bottom=2cm,left=3cm,right=3cm,marginparwidth=1.75cm]{geometry}
\usepackage{amsmath}
\usepackage{amssymb}
\usepackage{graphicx}
\usepackage[colorlinks=true, allcolors=blue]{hyperref}
\usepackage{cleveref}
\usepackage{siunitx}
\usepackage{diagbox}
\usepackage[section]{placeins}
\usepackage{float}
\usepackage{multirow}
\usepackage[table]{xcolor}
\usepackage{cite}
\usepackage{lineno}

\usepackage{todonotes}
\usepackage{caption}
\usepackage{subcaption}
\usepackage{authblk}

\usepackage{color}
\definecolor{Blue}{rgb}{0.00, 0.00, 1.00}
\definecolor{Red}{rgb}{1.00, 0.00, 0.00}

\title{Comparing the influence of Atlantic Multidecadal Variability and spring soil moisture on
European summer heat waves}

\author[1,+,*]{Valeria Mascolo}
\author[1,2,*]{ Cl{\'e}ment Le Priol}
\author[2]{Fabio D'Andrea}
\author[2]{Freddy Bouchet}

\affil[1]{Laboratoire de Physique à l'ENS de Lyon, CNRS, F-69342 Lyon, France}
\affil[2]{Laboratoire de Météorologie Dynamique, IPSL, ENS-PSL, CNRS, Paris, France}

\affil[+]{Corresponding author: valeria.mascolo@ens-lyon.fr}
\affil[*]{Authors contributed equally to the article.}
\date{ }
\usepackage{fancyhdr}

\fancyhfoffset[L]{1cm} 
\fancyhfoffset[R]{0cm} 
\begin{document}

\maketitle

\begin{abstract}

    
        In this work, we study and compare the influence of the Atlantic Multidecadal Variability (AMV) and spring soil moisture in Southern Europe on the duration and intensity of European summer heat waves.
        We study heat waves with return times of a few years and also propose a new methodological approach, \textit{return time maps}, that allows us to study rare heat waves with return times of 10 or 50 years.
    
        We use the outputs from three climate models, namely IPSL-CM6A-LR, EC-Earth3, and CNRM-CM6-1, in which North Atlantic sea surface temperatures are restored towards the observed AMV anomalies. 
        The three models give consistent results, with the exception of EC-Earth simulating a much greater influence of soil moisture.
    
        Our main conclusion is that spring soil moisture in Southern Europe is a slow driver of greater importance than the AMV for European summer heat waves, both in terms of the extension of the region of influence and in terms of amplitude.
        While the influence of the AMV concentrates over the very south of Europe, around the Mediterranean Basin, spring soil moisture influence extends over large parts of Europe.
        As might be expected, a positive AMV phase or low soil moisture generally induces hotter and longer heat waves.
        
        For more extreme events, the influences of the AMV and soil moisture increase, according to regional patterns that seem to be the same as for typical heat waves.
        However, confirming this statement would require datasets with more extreme events.
    
        \paragraph{Lay Summary}
    
    
        Beyond the daily fluctuations of the weather, the duration and intensity of heat waves can be modulated by slow drivers. In this work, we study and compare the influence of two slow drivers on the duration and intensity of summer heat waves in Europe. The first driver is a slow mode of variability of the North Atlantic Ocean sea surface temperature called the Atlantic Multidecadal Variability (AMV). The second one is the quantity of water available in the soil in spring in Southern Europe.
    
        We study typical heat waves that occur almost every year, but we also introduce a new method to study rare heat waves that occur only every 10 or 50 years, on average.
    
        Using results from global climate model experiments, we find that a positive AMV phase or low soil moisture generally induces hotter and longer heat waves, as could be expected.
        Our main result is that soil moisture is a slow driver of greater importance than the AMV. Indeed, its influence extends over a larger part of Europe and has more amplitude. 
        
    
    \end{abstract}

\section{Introduction}
\label{sec:intro}

    In a changing climate, extreme hot events are becoming more frequent and intense \cite{hansen_perception_2012, dunn2020, seneviratne2021}.
    The impacts of these events are detrimental on many levels, causing damage to our society, the environment, and other living beings \cite{portner2023}. 
    Europe, and especially the Mediterranean Basin, is identified as a hot spot for extreme hot events, with magnitudes changing according to future climate scenarios (see \cite{masson-delmotte2023} and references therein).
    In particular, heat extremes in Western Europe have warmed at a faster pace than elsewhere in the mid-latitudes \cite{rousi_accelerated_2022, vautard_heat_2023}.

    Besides global warming, different physical drivers influence the formation of heat waves, acting on different timescales~\cite{perkins_review_2015,horton_review_2016}.
    The fastest driver is the atmospheric circulation, which can cause heat waves through persistent high-pressure synoptic systems. The associated time scale corresponds to a few days, which is the typical duration of a heat wave.
    Slower drivers modulate the occurrence and frequency of heat waves, acting on seasonal, yearly, and multidecadal time scales \cite{perkins_review_2015}.

    Regarding summer European heat waves, an important seasonal driver is soil moisture. A soil moisture deficit in the Mediterranean Basin at the beginning of summer has been shown to act as a precondition for some extreme events such as droughts \cite{vautard_summertime_2007, zampieri_hot_2009} and heat waves \cite{fischer_contribution_2007, fischer_soil_2007, alexander_extreme_2011, materia_summer_2022} over continental Europe. 
    The mechanism is as follows: dry and warm air masses form over the dry soils of the Mediterranean and induce diminished cloudiness. These air masses are advected northward by southerly wind episodes, increasing temperature and evaporative demand over Europe, which in turn leads to drier soils. These drier soils amplify the warming through higher sensible heat emissions and favored upper-air anticyclonic circulation \cite{zampieri_hot_2009}.

    Sea surface temperature (SST) anomalies are another slow driver of heat waves, acting on different timescales. At the seasonal timescale, SST anomalies in the North Atlantic can favor heat waves through their influence on large-scale atmospheric modes like the NAO \cite{feudale2011, duchez2016, wulff2017, beobide-arsuaga2023a}.
    At the multidecadal timescale, SST anomalies are modulated by an internally-driven low-frequency mode of variability known as the Atlantic Multidecadal Variability (AMV).
    The AMV has been shown to influence the duration of heat waves in Europe \cite{sutton_atlantic_2012, qasmi_teleconnection_2017, qasmi_modulation_2021} 
    and to play a role in the occurrence of other extremes, such as droughts and precipitations, in other parts of the globe \cite{sutton_atlantic_2005, ruprich-robert_assessing_2017, ruprich-robert_impacts_2018}.
    

    While previous work has examined the influence of each driver separately, estimating the relative influence of each driver remains an open question. In this paper, we chose to compare the influence of two of these main drivers, namely spring soil moisture and the AMV, on the duration and intensity of European summer heat waves.
    
    A second question concerns the rarity of the events studied. Heat waves studied in previous work on the AMV \cite{ruprich-robert_impacts_2018, qasmi_modulation_2021} actually occur every one or two years. Therefore, they are not such rare events. However, the most harmful events are the largest and rarest ones \cite{robine_death_2008}. This calls for focusing on rarer, more extreme events. 
    In this paper, we propose a new tool, \textit{return time maps}, to study the influence of the AMV and spring soil moisture on rare European summer heat waves with return times of a few decades.

    We use two definitions of heat waves, which are complementary. The first one measures the number of heat wave days per year, based on the exceedance of two temperature thresholds \cite{lau_model_2012,qasmi_modulation_2021}.
    In the second definition, we fix the duration of the events studied and measure their intensity, characterized by the mean temperature anomaly during the event.

    The paper is organized as follows. In \cref{sec:dataandmethods}, we present the data used for this study and the two definitions of heat waves that we use. 
    In \cref{sec:AMV-sm-typical}, we compare the effects of the AMV and spring soil moisture on heat waves with return times of a few years.
    In \cref{sec:AMV-sm_rare}, we introduce a methodology to study rarer events with return times of a few decades and assess which driver, between the AMV and spring soil moisture, has the strongest influence on these rare heat waves. We summarize our findings in \cref{sec:SummaryOfResults} and discuss them in a broader perspective in \cref{sec:discussion}.

\section{Data and methods}
\label{sec:dataandmethods}
In this section, we present the data used for this study and the methodology to compare the influence of the AMV and spring soil moisture on heat waves independently. 
We then introduce the two different heat wave definitions that we consider in this study.

    \subsection{Data}
    \subsubsection{DCPP-AMV experiments}
\begin{figure}
    \centering
    \includegraphics[width=0.6\linewidth]{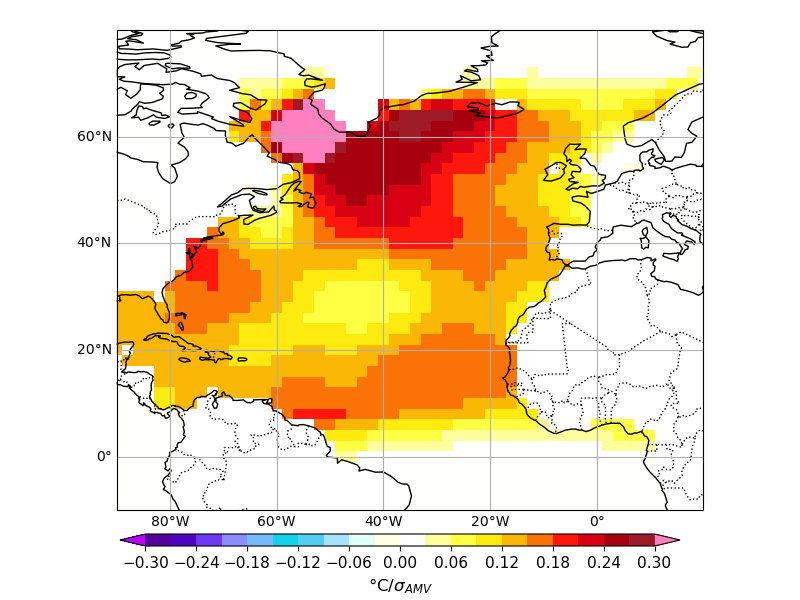}
    \caption{SST anomaly pattern used for restoring the SST in the DCPP-AMV experiments. Pattern courtesy of C. Cassou. \label{fig:AMV_pattern}}
    \end{figure}

    Given the relevance of the effects of the AMV on the climate and the potential predictability associated with it, there is ongoing work to deepen the understanding of its dynamical drivers and to improve its representation in models. One source of uncertainty regarding the impact of the AMV is the lack of a full understanding of the phenomenon itself. Additionally, model biases in representing crucial quantities for the AMV, such as the AMOC and teleconnection patterns, contribute to increasing this uncertainty. In this line of thought, it was shown in \cite{qasmi_teleconnection_2017} that CMIP5 models underestimate the ocean-atmosphere coupling at low frequency.
    The importance of dedicated modeling protocols to study decadal variability, such as the AMV, both at a global and regional scale using a coordinated multi-model approach has also been emphasized \cite{cassou_decadal_2018}.
    To this end, the Decadal Climate Prediction Project (DCPP), part of the CMIP6 project, was established \cite{boer_decadal_2016}.
    Within the DCPP, ensembles of simulations have been conducted to understand the predictability, variability, and impacts of decadal modes of climate variability such as the AMV.
    We use the outputs of the DCPP-C.1 experiments designed to enhance the understanding of the impact of the AMV on the global climate.
    In these experiments, the sea surface temperature (SST) of the North Atlantic is restored towards an anomalous SST pattern representative of the observed AMV, shown in \cref{fig:AMV_pattern}.
    The detailed procedure of the experiments can be found in the technical note of \cite{boer_decadal_2016}.

    The outputs of four coupled models that took part in the experiments were available on the Earth System Grid Federation. For one of them, HadGEM3-GC31-MM, the soil moisture outputs were not available, and therefore, we could not use it in our study. 
    We used the remaining three models: IPSL-CM6A-LR \cite{boucher_presentation_2020}, EC-Earth3 \cite{doscher_ec-earth3_2021}, and CNRM-CM6-1 \cite{voldoire_evaluation_2019} (hereafter simply referred to as IPSL, EC-Earth, and CNRM). 
    Further details about the models can be found in \cref{sec:supmat_method} and in the corresponding references.
    For each model, AMV+ and AMV- ensembles consisting of many 10-year members have been computed. 
    In the AMV+ ensemble, the SST is restored towards a positive anomaly pattern, while in the AMV- ensemble, it is restored towards its opposite. 
    For IPSL and CNRM, we also have a control run (CTRL) where the SST is nudged towards the climatology.
    The radiative forcing is set to its 1850 value. 
    Table~\ref{tab: Ensemble_sizes} summarizes the number of years available in each ensemble for each model.



    \begin{figure}
    \centering
    \includegraphics[width=0.7\linewidth]{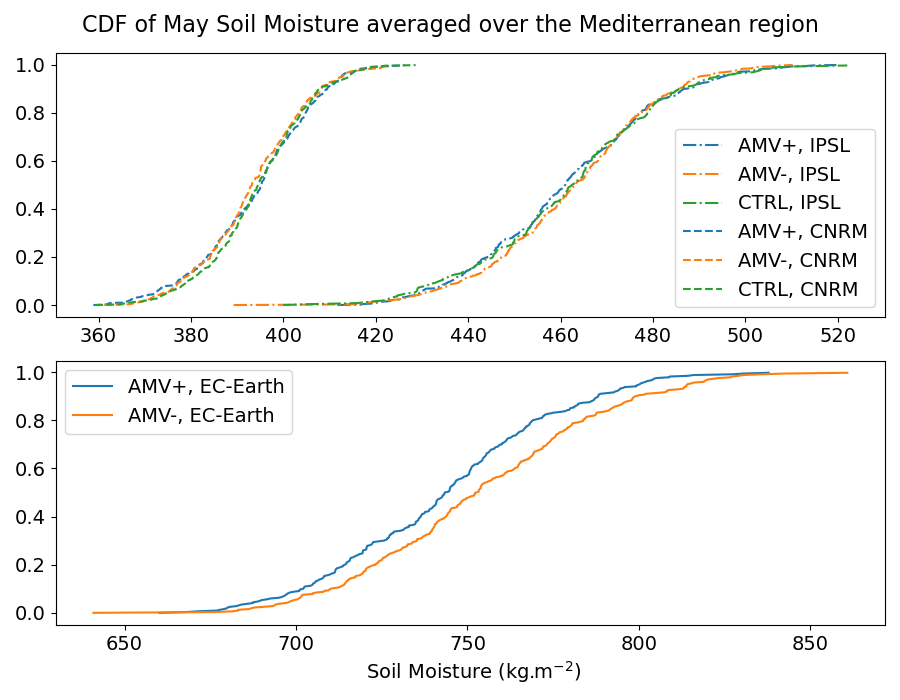}
    \caption{Empirical cumulative density function of May soil moisture averaged over the Mediterranean region (35°N-46°N, 10°W-30°E) in each model (IPSL, CNRM, EC-Earth) and each experiment (AMV+, AMV-, CTRL).\label{fig:mrso_CDF}}
    \end{figure}

    \begin{center}
    \begin{table*}
    \centering
    	\begin{tabular}{c | c c c}
            \diagbox{Dataset}{Model} & IPSL & EC-Earth & CNRM  \\ \hline
    		
    		AMV+/- & 500 & 320 & 390/400  \\ \hline
    		CTRL & 250 & 0 & 400  \\ \hline
    		Dry/Wet & 500 & 250 & 450 \\
    		\hline
    	\end{tabular}
    	\caption{\textit{Number of years in all ensembles for each model}. The AMV+/- and CTRL ensembles consist of simulation members of 10 years. The Dry and Wet ensembles are built by sorting the years in all simulation members according to their average May Soil Moisture value over Southern Europe (see \cref{sec:Dry and Wet ensembles,sec:supmat_method} for details). 
    	For CNRM, some surface air temperature and soil moisture outputs are missing. 
    	 \label{tab: Ensemble_sizes}}
    \end{table*}
    \end{center}

    \subsubsection{Observation-derived datasets for soil moisture}

    We use two observational datasets of land variables, ERA5-Land and GLEAM v3.8a, as benchmarks against which to compare the persistence of soil moisture anomalies in the three models.

    ERA5-Land is a reanalysis dataset of land variables that describes the evolution of the water and energy cycles over land in a consistent manner~\cite{munoz-sabater_era5-land_2021}. It goes back to 1950 and is produced through global high-resolution numerical integrations of the ECMWF land surface model driven by the downscaled meteorological forcing from the ERA5 climate reanalysis.
    We use monthly averages of the volumetric soil water in the top 3 (resp. 4) soil layers of the model, corresponding to depth 0-100 (resp. 0-289) cm. 

    The Global Land Evaporation Amsterdam Model (GLEAM) is a set of algorithms dedicated to the estimation of terrestrial evaporation and root-zone soil moisture from satellite data \cite{miralles_global_2011,martens_gleam_2017-1}. The surface soil moisture is assimilated from satellite microwave remote sensing data. The model features a multi-layer soil model driven by satellite observations of precipitation with fast and slow draining of water from the surface layer towards the deepest layers. 
    The soil moisture estimates are validated against 2325 soil moisture sensors across a broad range of ecosystems.
    We use GLEAM v3.8a monthly average datasets of the surface (0-10 cm depth) and root zone (10-100 cm) soil moisture. We perform a weighted average of the two datasets to obtain the soil moisture between 0 and 100 cm. GLEAM datasets run from 1980 to the present.
     
    \subsection{Influence of the AMV on spring soil moisture and creation of the Dry and Wet ensembles}
    \label{sec:Dry and Wet ensembles}
    To assess the influence of spring soil moisture on summer heat waves, we use the outputs of the three experiments (AMV+/- and CTRL) for each model to build two new ensembles, called Dry and Wet, corresponding to low and high spring soil moisture, respectively.
    We use the monthly total soil moisture content (mrso), which is the only relevant variable available for all three models.
    First, we compute the average May soil moisture $\text{SM}_{\text{av}}$ for each year and each experiment over a domain covering Southern Europe. Following \cite{vautard_summertime_2007}, we choose the extent of the domain to be the rectangular box 35°N-46°N, 10°W-30°E. 
    Then, we verify the influence of the AMV on $\text{SM}_{\text{av}}$ in the models by plotting the cumulative density function (CDF) of $\text{SM}_{\text{av}}$ for each experiment in \cref{fig:mrso_CDF}.
    The influence of a positive versus negative phase of the AMV on $\text{SM}_{\text{av}}$ is negligible in IPSL and CNRM ($-0.8$ and $+0.6$ kg.m$^{-2}$ respectively, to be compared to standard deviations of 19.8 and 11.8 kg.m$^{-2}$) and relatively small in EC-Earth ($-10.5$ kg.m$^{-2}$ for a standard deviation of 34.4 kg.m$^{-2}$).
    
    In order to rigorously compare the influence of spring soil moisture with the one of the AMV, we must avoid an imbalance of years from the AMV+ and AMV- experiments in the Dry and Wet ensembles. Indeed, if the Dry ensemble were to contain 200 years from the AMV+ experiment versus 100 from the AMV- experiment, and vice versa for the Wet ensemble, we would observe a partial influence of the AMV when comparing the Dry and Wet ensembles. 
    To ensure that there is no indirect influence of the AMV, we impose the constraint that an equal number of years come from the AMV+ and AMV- experiments in the Dry and Wet ensembles\footnote{This constraint also helps address the imbalanced size of the AMV+/- ensembles of the CNRM model. Indeed, 120 years of soil moisture outputs are missing for the AMV- experiment with this model.}. 
    We then place years with low (resp. high) $\text{SM}_{\text{av}}$ in the Dry (resp. Wet) ensemble. We refer the reader to \cref{sec:supmat_method} for more details about the procedure and \cref{fig:Dry/wet_construction} for an illustration.
    The mean value of the soil moisture inside the Dry (resp. Wet) ensemble is nearly one standard deviation below (resp. above) the mean value of the soil moisture averaged over the AMV+/- and CTRL experiments all-together.
    This makes the comparison with the AMV experiments meaningful because the targeted SST pattern for the relaxation corresponds to one standard deviation of the AMV variability: we compare the influence of a one sigma anomaly for both the AMV and spring soil moisture.

    \subsection{Heat wave definitions}

    \subsubsection{Threshold-based definition}
    \label{def:hw-threshold}
 
    We call the first definition that we use \textit{threshold-based definition} as it relies on two temperature thresholds. 
    It was first introduced in \cite{lau_model_2012} and has been used in \cite{qasmi_modulation_2021} and \cite{ruprich-robert_impacts_2018} to study the influence of the AMV on heat waves in Europe and North America respectively.
    Using this definition enables us to compare our results with \cite{qasmi_modulation_2021}, which used an earlier version of the CNRM and EC-Earth models.
    According to this definition, a heat wave event is a group of days that satisfies the following three conditions: 
    \begin{enumerate}
        \item[(i)] $T_{\textrm{max}}$ must exceed $T_1$ for at least 3 consecutive days,
        \item[(ii)] $T_{\textrm{max}}$ averaged over the entire event must exceed $T_1$ and
        \item[(iii)] $T_{\textrm{max}}$ on each day of the event must exceed $T_2$.
    \end{enumerate}
    where $ T_{\textrm{max}}$ is the daily maximum 2-meter air temperature and $T_1$ and $T_2$ are two temperature thresholds corresponding respectively to the 90th and 75th percentile of the local June-July-August (JJA) $T_{\textrm{max}}$ distribution built from the $T_{\textrm{max}}$ values of all members of the AMV+ and AMV- simulations (for each model). 
    This definition is location-dependent, since the $T_{\textrm{max}}$ distribution varies with latitude and longitude.
    For each grid point, we count the number of heat wave days in each year.
    In this study, we are interested in the response, in terms of heat wave days per year, to an AMV-forcing. 
    In \cref{sec:AMV-sm-typical}, we consider the mean difference between AMV+ and AMV- for each model, 
    and in \cref{sec:AMV-sm_rare}, we condition on 
    the most extreme years, i.e., the ones with the highest number of heat wave days. 
    We make the choice of defining a heat wave by the absolute temperature during the summer and not on the location-dependent anomaly. Our choice is more relevant for those health impacts which are caused by the absolute temperature and not the anomaly. On the other hand, the alternative choice  of using use the temperature anomaly might be more suited for some ecosystems or agricultural impacts.
    
    \subsubsection{14-day heat waves}
    \label{def:hw-14days}

    To quantify the heat wave intensities for several independent durations, heat wave indices based on the combined temporal and spatial averages of the surface or 2-meter air temperature have been adopted in many studies. Notably, seminal studies of the 2003 Western European and 2010 Russian heat waves considered the averaged temperature over variable long time periods (7 days, 15 days, 1 month, and 3 months) \cite{schar_role_2004,barriopedro_hot_2011,coumou_decade_2012}. Similar definitions have been adopted in a set of recent studies \cite{ragone_computation_2018,ragone_computation_2020,ragone_rare_2021,galfi_large_2019,galfi_fingerprinting_2021,galfi_applications_2021}. This viewpoint is expected to be complementary with more classical heat wave definitions \cite{perkins_review_2015} and extremely relevant to events with the most severe impacts.
    For our second definition, we consider averages of the daily maximal 2-meter air temperature over a period of 14 days. This duration corresponds to long-lasting events that are more impactful than events lasting only a few days, while still being much shorter than the seasonal time scale. 
    Mathematically, the definition reads as follows:
    \begin{equation}\label{eq: HW_def2}
    \Tilde{A}(\mathbf{r}) = \max_{t, t+D \in \text{JJA}} \frac{1}{D} \int_t^{t+D} \left(T_{\textrm{max}}(\mathbf{r}, t') - \bar{T}_{\textrm{max}}(\mathbf{r}) \right) \mathrm{d}t' \, ,
    \end{equation}
    where $\mathbf{r}$ and $t'$ represent the spatial and time coordinates respectively, and $D=14$ days is the heat wave duration.
    We do not perform any spatial averaging because we aim to maintain a local perspective, as with the threshold-based definition.
    To facilitate the comparison across different locations, we subtract $\bar{T}_{\textrm{max}}(\mathbf{r})$, the local June-July-August (JJA) average of $T_{\textrm{max}}(\mathbf{r}, t)$. 
    This allows us to measure a temperature anomaly relative to the local seasonal mean. 
    We compute one value of $\tilde{A}$ for each year and at each grid point.

\section{Influence of the Atlantic Multidecadal Variability and spring soil moisture on heat waves with return times of a few years}
\label{sec:AMV-sm-typical}

In this section, we compare the effect of the AMV and spring soil moisture on typical heat waves with return times of one or two years. Following previous studies \cite{ruprich-robert_impacts_2018, qasmi_modulation_2021}, we first 
use the threshold-based definition introduced in \cref{def:hw-threshold} to measure the drivers' influence on the mean number of heat wave days per year. 
We then use the second definition, presented in \cref{def:hw-14days}, to measure their influence on the 14-day heat wave intensity.

    \subsection{Influence on the frequency of heat wave days}\label{sec:influence typical HW def1}
    
\begin{figure}[bth]
    \centering
    \includegraphics[width=1.05\textwidth]{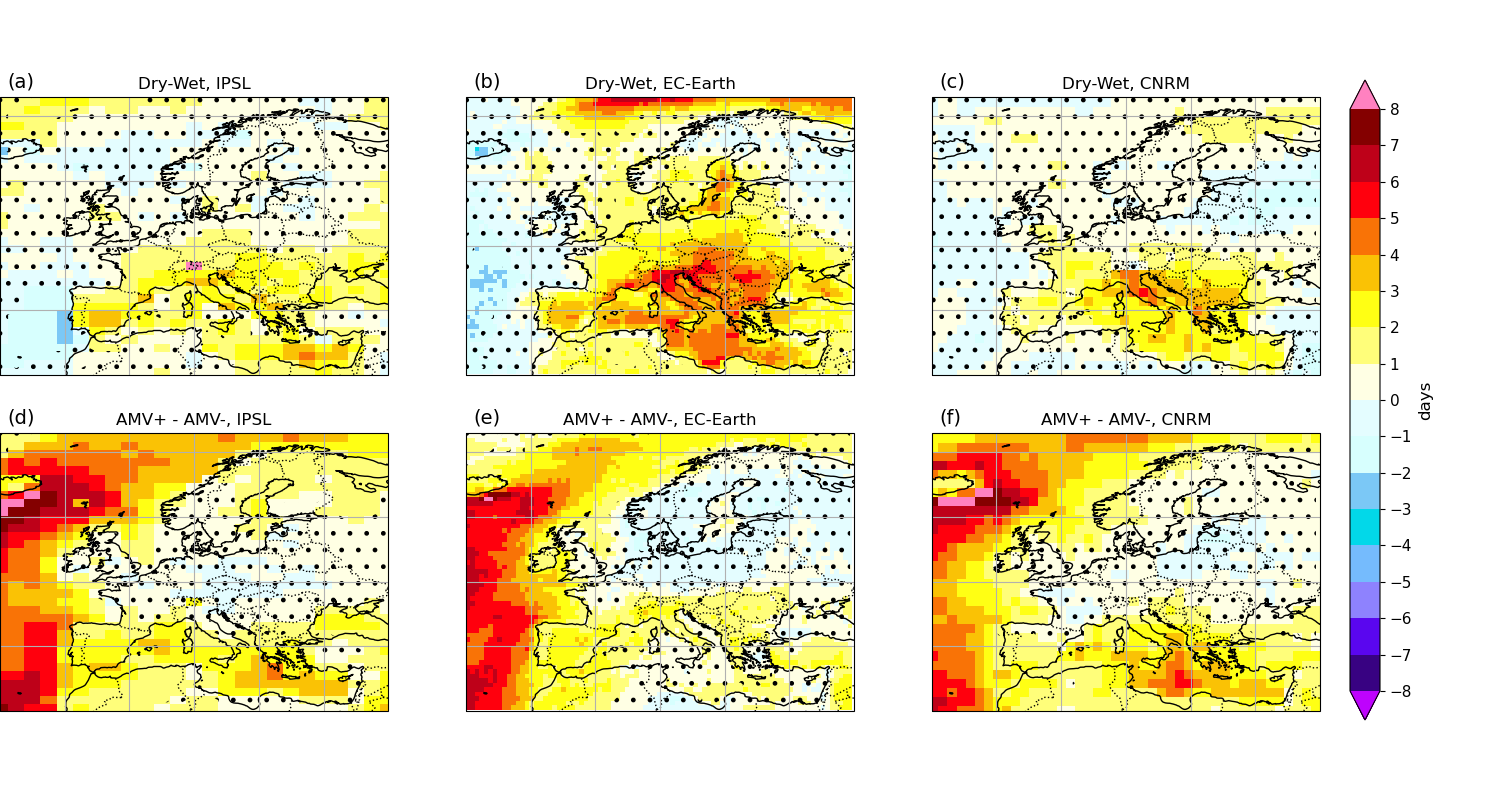}
    \caption{\label{fig:mean_anomaly_RR} \textit{Mean response maps for the frequency of heat wave days.}
    Differences in the mean number of heat wave days per year between the Dry and the Wet ensembles (top line) and the AMV+ and AMV- ensembles (bottom line) for each of the three models. Stippling denotes area below the 95\% significance level according to a bootstrap test.  }
    \end{figure}

    We first study the effect of the AMV and spring soil moisture on heat waves using the threshold-based definition introduced in \cref{def:hw-threshold}.
    \Cref{fig:mean_anomaly_RR} shows the difference in the mean number of heat wave days per year between the Dry and Wet ensembles  and the AMV+ and AMV- ensembles  for each of the three models. 
    We use a bootstrap to test the significance of the differences at the 95\% level (see \cref{sec:supmat_method} for details).
    Both the AMV and spring soil moisture influence significantly the mean number of heat wave days over some areas of Europe but according to different regional patterns.

    For the AMV, all models agree on a significant positive influence on southern Europe (Iberian Peninsula, Italy, and Greece) around the Mediterranean Basin with positive differences ranging from 1 to 3 heat wave days per year. 
    These numbers must be compared with the mean number of heat wave days per year in the AMV- ensemble, which ranges between 7 and 10 heat wave days per year in most locations (not shown). In EC-Earth, the region of positive influence extends northwards up to approximately 48° North.
    Differences over the rest of the continent range between -1 and +1 heat wave day per year and do not pass the significance test.
    We note that all models show a region of small negative influence around Poland (between 0 and -1 heat wave days per year). Although this difference does not pass the test, the consistency between the three models suggests the potential presence of a real signal.
    For the current heatwave definition (threshold-based) the strongest response, in terms of amplitude, is located over the North Atlantic ocean, with all the models giving a consistent response of a clear anomaly in heatwaves days for AMV+ phase. We will see in the following section that this is not the case for the other definition. We leave this point for further analysis, as we rather focus over land heatwaves.  

    Regarding the influence of soil moisture, the models present less consistent results. 
    On one hand, CNRM and IPSL show a positive response over Southern Europe, mostly ranging between 1 and 3 heat wave days per year, with some locations reaching up to 4 heat wave days per year. In these two models, the influence of soil moisture extends further north than the one of the AMV and is slightly larger but of comparable amplitude. 
    On the other hand, EC-Earth simulates a much more extended and much larger positive influence of spring soil moisture deficit. The region of positive difference covers almost all of Europe as well as Northern Africa, with differences ranging from 3 to 6 heat wave days per year over Spain, Italy, and Southern Central Europe.
    In this model, spring soil moisture has thus a much stronger influence than the AMV. 
    Once again, regions where the differences are between -1 and +1 heat wave day per year do not pass the significance test.
    
    \begin{figure}
        \centering
        \includegraphics[width=0.6\textwidth]{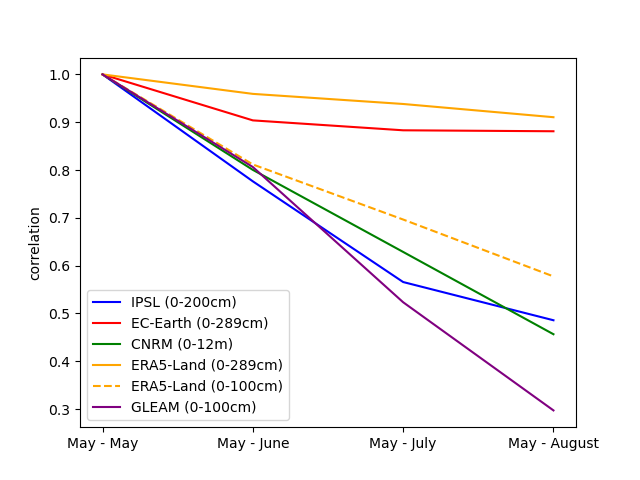}
        \caption{Pearson correlation coefficients between the month of May and the months of June, July, and August of the monthly averaged  soil moisture averaged over Southern Europe (10°W-30°E,35°N-46°N).}
        \label{fig:correlation-mrso}
    \end{figure}

    \paragraph{Discrepancy between EC-Earth and the two other models.}
    EC-Earth presents a response to spring soil moisture deficit that is both more extended and stronger than in the other two models.
    This stronger response is associated with a strong positive Zg500 anomaly and reduced precipitation over Central Europe (not shown).
    In order to further investigate the cause of this discrepancy, we show in \cref{fig:correlation-mrso} the auto-correlation function of the monthly averaged soil moisture averaged over Southern Europe (10°W-30°E,35°N-46°N).

    
    In IPSL and CNRM, the auto-correlation decays almost linearly from 1 for the May-May correlation to about 0.5 for May-August. On the other hand, in EC-Earth, the correlation levels out for the three summer months and is nearly constant at about 0.9\footnote{The different auto-correlations between models may be due to different land-atmosphere couplings or to intrinsic differences between land models. Among the latter, a notable difference is the difference in soil depth relevant for the total soil moisture content. One may wonder if this depth difference dominates the auto-correlation difference between models. It appears not to be the case, as IPSL and CNRM share similar auto-correlation decays despite having soil depths of 2 and 12 meters respectively.}.
    This larger correlation implies a stronger persistence of soil moisture anomalies in EC-Earth: dry soils in May remain abnormally dry throughout the summer, causing the surface air to heat up more strongly due to a larger sensible heat flux.
    For comparison, we also plot the auto-correlation of the soil moisture in the observation-derived datasets ERA5-Land and GLEAM v3.8a.
    ERA5-Land shares the same land model, HTESSEL, with EC-Earth and also presents a strongly persistent auto-correlation of soil moisture. This indicates that the large correlation of EC-Earth is not due to an abnormal atmospheric response but rather to the land model itself.
    In an attempt to estimate if HTESSEL produces too large auto-correlation of soil moisture, we compare the auto-correlation of the 1st meter of soil with GLEAM v3.8a, a product more directly derived from satellite observations than ERA5-Land. 
    The auto-correlation for the months of July and August remains larger in ERA5-Land compared to GLEAM. Thus, it could be that HTESSEL produces too large persistence of soil moisture anomalies. However, it would be hasty to draw any definitive conclusions. While we consider the results of EC-Earth with some caution, we cannot rule it out.
    
    \subsection{Influence on the intensity of 14-day heat waves}
    \label{sec:influence typical HW def2}
    
\begin{figure}[bth]
    \centering
    \includegraphics[width=1.15\textwidth]{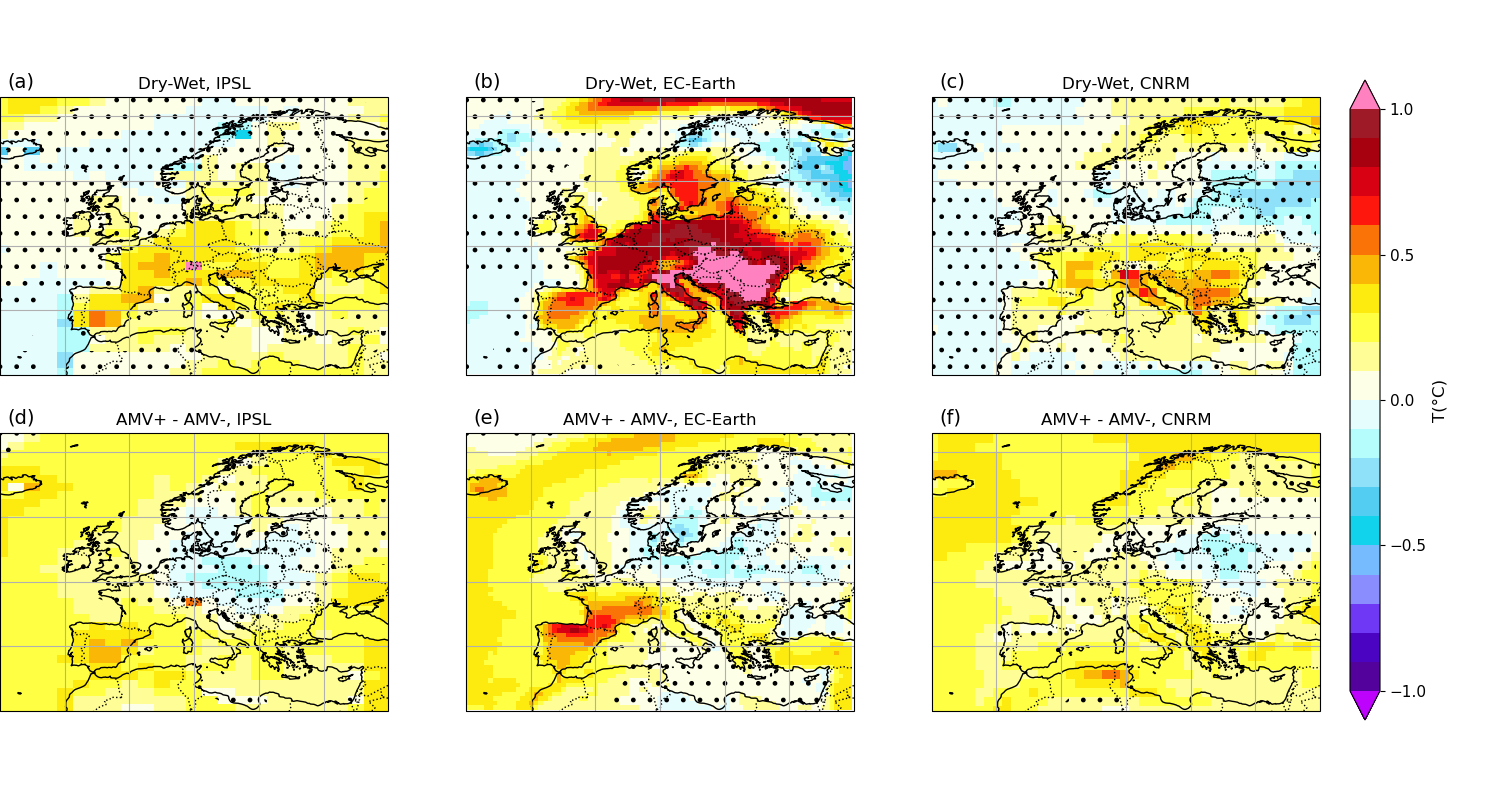}
    \caption{\label{fig:mean_anomaly_Max14dAv} \textit{Mean response maps for 14-day heat waves.} Difference in the mean value of the 14-day averaged temperature $\tilde{A}$ (defined in \cref{eq: HW_def2}) between (top line) the Dry and the Wet ensembles and (bottom line) the AMV+ and AMV- ensembles for each of the three models. Stippling denotes area below the 95\% significance level according to a bootstrap test.}
    \end{figure}

    We turn to the second definition to assess the influence of the AMV and spring soil moisture on the intensity of 14-day heat waves.    
    \Cref{fig:mean_anomaly_Max14dAv} shows the mean difference in the 14-day averaged temperature $\tilde{A}$ (see \cref{def:hw-14days}) between the Dry and Wet (top line) and AMV+ and AMV- (bottom line) ensembles.
    Note that these maps should not be interpreted as the temperature averaged over any given period of the summer. Indeed, the 14 days that contribute to the average are a priori different for each grid point and for each year of the simulations.
    The patterns of influence are very similar to those observed in \cref{fig:mean_anomaly_RR} for heat wave duration.

    \vspace{0.5cm}

    All three models agree on a positive influence of the AMV that concentrates around the Mediterranean Basin, where differences range from 0.2 to 0.4°C in most places. 
    EC-Earth simulates higher differences, ranging from 0.5 to 0.8°C, in Northern Spain and Southern France. 
    All models simulate a positive influence over Turkey, and this extends around the Black Sea for IPSL. 
    CNRM and IPSL also simulate positive differences over parts of Scandinavia, but they do not agree on the localization. 
    As for the first definition, all models simulate a region of small negative influence over Poland and neighboring countries that do not pass the significance test but may be a hint of a signal.

    The influence of spring soil moisture extends over larger areas of Europe.
    For CNRM, the region of positive influence extends from the Atlantic Ocean to the Black Sea below 50°N. For IPSL and EC-Earth, it extends over almost all continental Europe. Regions of exception are Scandinavia, the Baltic countries, Poland, and Russia in IPSL, and Northern Scandinavia and Russia in EC-Earth. 
    The influence of soil moisture is also larger than that of the AMV, with differences reaching up to 0.5 or 0.6°C in many locations for IPSL and CNRM.
    As for heat wave duration, EC-Earth simulates a much larger influence of low soil moisture with positive differences above 0.8°C over all Western and Central Europe and peak differences up to 1.3°C over the Balkans.   
    As for the first definition, we expect the stronger Dry-Wet response of EC-Earth to be due to the larger auto-correlation of soil moisture.
    \vspace{0.5cm}

    Combining the results of sections \ref{sec:influence typical HW def1} and \ref{sec:influence typical HW def2}, we conclude that spring soil moisture 
    in Southern Europe is a slow driver of greater importance than the AMV for European summer heat waves: its region of influence covers larger areas of Europe with a larger amplitude.
    We note that EC-Earth presents a different result than the other two models, with a much stronger influence of spring soil moisture. Regarding this point, we do not have enough evidence to conclude that one model is more or less biased than the others. 
    We also tried to investigate the mechanisms by which the AMV influence the frequency and intensity of heat waves over Europe. 
    To this aim, we plotted the mean response maps to the AMV phase for cloud cover, latent heat flux, sensible heat flux, 500hPa geopotential height, precipitation, and soil moisture averaged over the June, July and August period in \cref{fig:all_fields_JJA_response} (see \cref{sec:supmat_figures}). 
    There is no significance in the responses for individual models. 
    This lack of significance does not allow us to address the important question of the mechanism by which the AMV influences heatwaves with these datasets. This calls for either longer datasets or the application of stronger SST forcing in the numerical experiments as in \cite{qasmi_modulation_2021}, where the authors simulate the influence of one-sigma, two-sigma, and three-sigma AMV anomalies. 
    The experiments with stronger forcing yield significant signals, allowing the authors to conclude that a positive AMV phase is associated with "drier soils and a reduction of cloud cover" around the Mediterranean Basin and "an enhancement of the downward radiative fluxes over lands" \cite{qasmi_modulation_2021}.

\section{Influence of the Atlantic Multidecadal Variability and spring soil moisture on heat waves with return times from 10 to 50 years}
\label{sec:AMV-sm_rare}

The heat waves that we considered in the previous section are not really rare events. 
Indeed, the second definition gives one value of $\tilde{A}$ for each year, while for the first definition the fraction of years during which at least one heat wave occurs is consistently comprised between 50\% and 80\% for all models and land grid points (\cref{fig:frac_HW_yearT9075}).
This means that such heat waves occur every one or two years and are thus not so rare events.
In this section, we focus on rarer events with return times of a few decades.
In \cref{sec:local_RT_plots}, we use return time plots to see how the influence of the slow drivers evolves with the return time for a single grid point. However, this provides information only at a local level.
To study extreme events at the European level, we introduce return time maps in \cref{sec:RT_maps}.
Based on these maps, we study extreme events with return times ranging from 10 to 50 years.

    \subsection{Study of rare events using local return time plots}
    \label{sec:local_RT_plots}

    \begin{figure}
    \centering
    \includegraphics[width=1\textwidth]{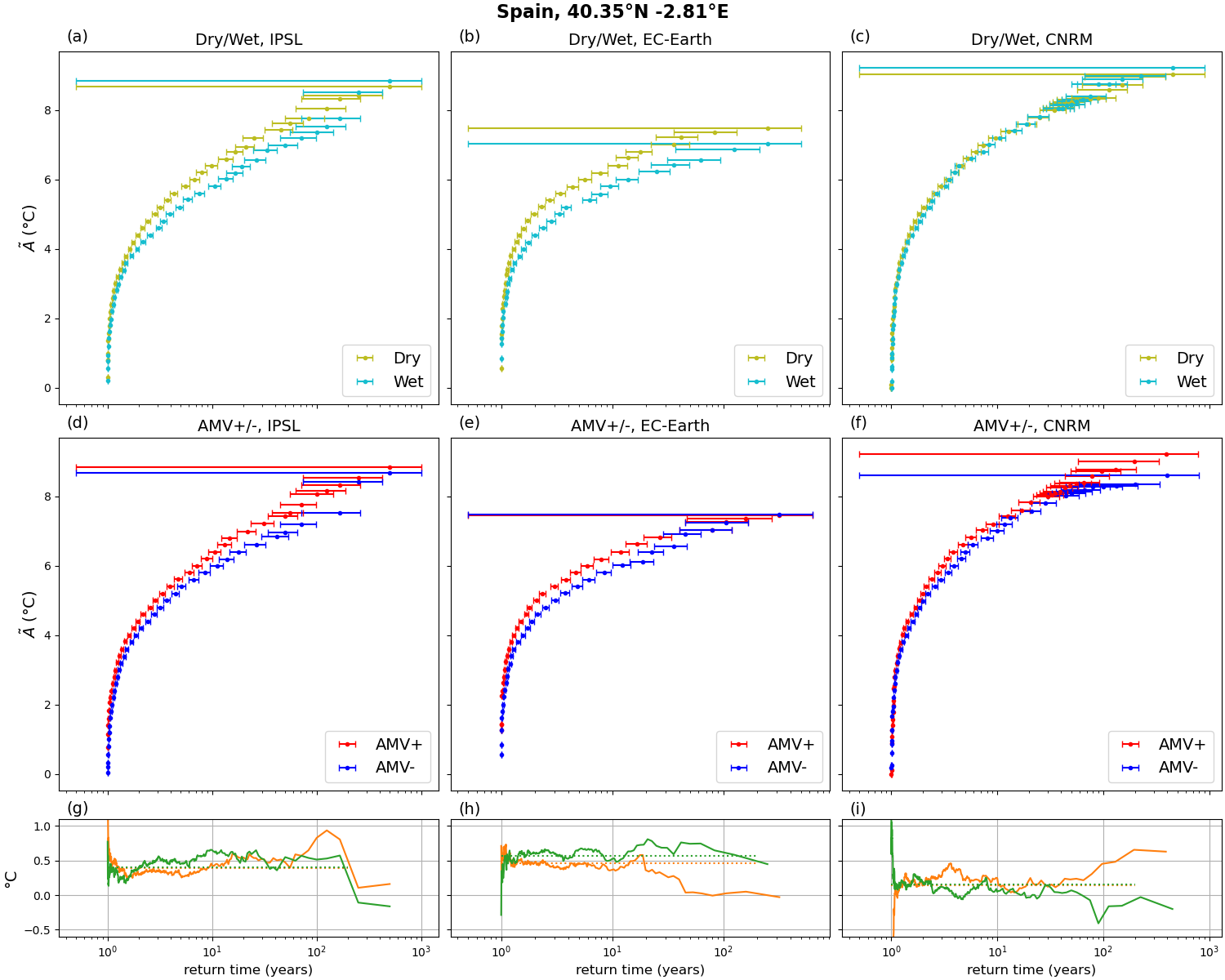}
    \caption{\label{fig:RT_Spain_Max14dAv} \textit{Return time curves of the intensity of 14-day heat waves}. The two first lines show the temperature anomaly as a function of the return time for the dry and wet ensembles (top row) and the AMV+ and AMV- ensembles (second row) for the three models. Errorbars correspond to one standard deviation of the estimated return time to observe an event of a given amplitude.
    The third row shows the Dry - Wet (green) and AMV+ - AMV- (orange) differences as a function of the return time. The dotted lines indicate the mean difference values.
    For each model the chosen grid point covers the coordinate indicated in the figure title which corresponds to a location in the Northeast of Spain, close to the Mediterranean Sea.}
    \end{figure}

    In climate statistics, the probability of an event is often expressed in terms of return time:
    if an event has a probability $1/Y$ to occur each year, then it has a return time of $Y$ years, 
    which also corresponds to the average duration between two such events.
    A classical way to visualize the intensity of events of decreasing probability
    is to build return time curves:
    events are ranked by decreasing intensity $a_1 > a_2 > ... > a_N$ and 
    the empirical return time associated to the level $a_m$ reads:
    \begin{equation}\label{eq: return time formula}
        \hat{r}(a_m) = \frac{N}{m} .
    \end{equation}
    This is simply the inverse of the empirical probability to have an event of intensity at least as large as $a_m$ which is $m/N$.
    By construction, the minimal event's intensity has a return time of one year,
    the median a return time of two years and the largest event is associated with a return time of $N$ years.
    In the present study, $a$ will be either the number of heat wave days in a year (following the definition of \cref{def:hw-threshold}) or the 14-day heat wave intensity defined in \cref{def:hw-14days}.
    Remember that both quantities are defined at each grid point.
    We present results for the second definition in this section. 
    The results for the first definition are similar and the corresponding figures are shown in \cref{sec:supmat_figures}.

    \Cref{fig:RT_Spain_Max14dAv} shows the return time curves for the intensity of 14-day heat waves for a grid point in the Northeast of Spain. 
    We chose this region for illustration purposes, as it is one where we observe a strong response of $\tilde{A}$ to both the AMV and the soil moisture in all models, except for the AMV in CNRM (\cref{fig:mean_anomaly_Max14dAv}).
    The six upper panels show the empirical return time curves for each ensemble (AMV+, AMV-, Dry, Wet) and each model\footnote{The curve for the dry/wet ensembles on one side and for the AMV+/- ensembles on the other side are very similar for each model because they are built from the same dataset.}. 
    Several considerations arise from these plots. 
    The temperature anomaly starts at about 0°C for a return time close to 1 year\footnote{Very few events have an anomaly between -2°C and 0°C. They correspond to extreme cold years. They have been cropped for the sake of plot's readability.}, 
    goes up quickly to around 4 or 5°C for a return time of 2 years and then rises at a slower pace, up to anomalies of 7°C to 9°C, depending on the model, for return times of a few centuries.
    More importantly, these curves  allow to visualize how the influence of the drivers
    on the frequency of events changes as the return level increases.
    For temperature anomalies from 4°C to 6°C a low soil moisture (resp. a positive AMV phase) double the frequency of occurrence with respect to a high soil moisture (resp. a negative AMV phase) in IPSL and EC-Earth.
    In CNRM the AMV response is smaller and there is no visible influence of soil moisture. 

    It is less straightforward to visualize the changes in the influence of the drivers on the intensity of the events. 
    At a first glance, it seems that the influence on the intensity increases with the return time. However, this interpretation is misleading, as it is influenced by the near-vertical alignment of the curves for the lowest return times.
    Indeed, if two curves are concave and a constant vertical distance away, then the horizontal distance between them grows along the x-abscissa\footnote{
    As a simple example, one can consider the two curves $y_1(x) = \sqrt{x}$ and $y_2(x)=y_1(x) + \Delta y$. The horizontal distance between them is $\Delta x = 2\Delta y \sqrt{x} - \Delta y ^2$ which grows as $\sqrt{x}$. 
    }.
    To properly visualize the evolution of the influence of the drivers as the return time increases, 
    we show the Dry - Wet and AMV+ - AMV- differences as a function of the return time on the third line of \cref{fig:RT_Spain_Max14dAv}.
    We see that the influence of both the AMV and soil moisture is roughly constant from a return time of less than two years up to a few decades. There are large fluctuations at very short and large return times which correspond to the two tails of the distribution.
    The influences of the AMV and spring soil moisture are of the same order within each model. However, they are lower in CNRM than in the two other models. \Cref{fig:RT_Spain_RR} shows similar return time plots for the threshold based definition.

    While return time plots give a view of the evolution of extreme events intensity when shifting from small to large return times, the information they provide is location dependent.
    For instance, while EC-Earth produces noticeably lower extreme heat waves than the other two models at the location examined in \cref{fig:RT_Spain_Max14dAv}, it turns out to be the opposite in Central Europe (not shown).
    We present in the next section a method to overcome this limitation and obtain a regional picture of the influence of the drivers on extreme events.

    \subsection{Regional picture of AMV and spring soil moisture influence on heat waves with return times of 10 and 50 years}\label{sec:RT_maps}

    In order to synthesize at the continental scale the local results described above, we build on the concept of return time curve to introduce \textit{return time maps}.
    To put it in a nutshell, these maps allow to visualize the difference in intensity of events with the same return time but belonging to different ensembles. As far as we know, we are the first to plot this kind of maps. A similar but different concept is Risk Ratio maps where the changes in probability of events of the same intensity are displayed~\cite{kharin2018}.
    We first explain how these maps are built before discussing the results.

    \paragraph{Method}
    For each grid point, we consider the difference between the AMV+ and AMV- or Dry and Wet return time curves averaged over all the events with a return time larger than a threshold RT. 
    We considered threshold values of RT=10 and 50 years, which provide a compromise between studying extreme events and keeping enough events to calculate statistics.

    To be more specific, for each dataset $\mathcal{S} \in$ \{AMV+, AMV-, Dry, Wet\} and for the return times RT=10 or 50 years, we compute at each grid point:
    \begin{equation}
        a^{\mathcal{S}}_{\textrm{RT}}(\mathbf{r}) = \frac{1}{K} \sum_{i=1}^K a_i(\mathbf{r}) \quad
    \text{where} \; K=\frac{N_{\mathcal{S}}}{\textrm{RT}}  
    \end{equation}
    where $\mathbf{r}$ is the spatial coordinate, $N_{\mathcal{S}}$ is the number of years available in the dataset $\mathcal{S}$ and $K$ is the number of years such that $\hat{r}(a_i) \geq \textrm{RT}$.
    We recall that the $a_i$'s are the number of heat wave days in a year or 14-day heat wave intensity, depending on the definition considered.
    We have assumed that $a_1>a_2>a_3>...>a_{N_{\mathcal{S}}}$ in each dataset.
    The interpretation is that $a^{\mathcal{S}}_{10yrs}$ is the empirical mean of 10-year events in the dataset $\mathcal{S}$.
    We then display for each grid point the differences
    $a^{\textrm{AMV+}}_{\textrm{RT}} (\mathbf{r}) - a^{\textrm{AMV-}}_{\textrm{RT}} (\mathbf{r})$ and
    $a^{\textrm{Dry}}_{\textrm{RT}} (\mathbf{r}) - a^{\textrm{Wet}}_{\textrm{RT}} (\mathbf{r})$.
    Note that these differences can be seen as computing the differences between the two curves in panels (a) to (f) of \cref{fig:RT_Spain_Max14dAv} (these differences are shown in panels (g) to (i)) and then taking the average of these differences over the points with return time greater than 10 or 50 years only, i.e. on the rightmost part of these curves.
    We call these maps \emph{return time maps}.
    \vspace{0.7cm}

    \begin{figure}
    \centering
    \includegraphics[width=1.05\textwidth]{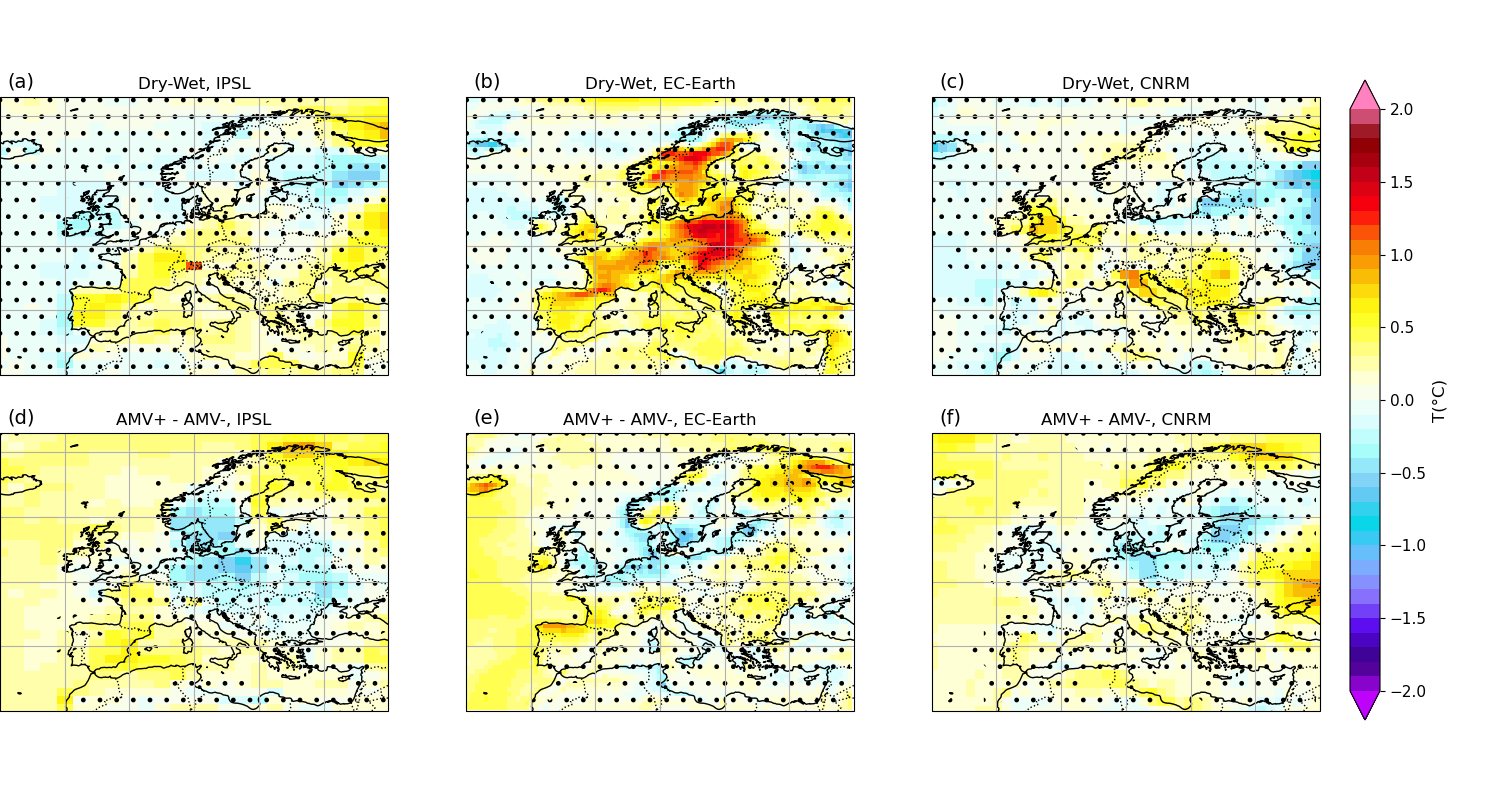}
    \caption{\label{fig:RT=10yrs_anomaly_Max14dAv} \textit{10 years return time maps for 14-day heat waves.} The maps show the  difference of $\tilde{A}$ between (top line) the Dry and Wet ensembles and (bottom line) the AMV+ and AMV- ensembles, conditioned over events with return time greater than 10 years. Stippling denotes area below the 95\% significance level according to a bootstrap test.}
    \end{figure}

    \Cref{fig:RT=10yrs_anomaly_Max14dAv} shows the 10-year return time maps for the intensity of 14-day heat waves. 
    Each individual map shows only reduced area of statistical significance compared to the mean response maps. This is because we consider only the most extremes years, which reduces the number of years included in the maps.
    Regarding the AMV influence, there is almost no region of significance.
    Larger regions of significance subsist for spring soil moisture: for IPSL, a region of positive influence extends over Western Europe, up to 52°N, while this positive influence covers almost all Europe until about 25°E in EC-Earth. 
    In many areas, the 10-year response is higher than the mean response with an amplitude up to 1°C. Once again, EC-Earth shows a higher response to soil moisture with a peak amplitude of 1.8°C. We note that the peak of this response has shifted northward with respect to the mean response.
    We note that the regional response patterns are similar to those of the mean response (see \cref{fig:mean_anomaly_Max14dAv}; note that the color scale is different). 
    Therefore, it seems that response patterns for the rarest events are similar to the patterns for less rare events. However, confirming this statement would require datasets with more extreme events.

    We show the 50-year return time maps of 14-day heat waves intensity in \cref{fig:RT=50yrs_anomaly_Max14dAv} in \cref{sec:supmat_figures}.
    These maps are much noisier than the 10-year return time maps. This is because these maps correspond to events that are in the very tail of the empirical distribution, where the fluctuations of the response are large, as can be seen on panels (g,h,i) of \cref{fig:RT_Spain_Max14dAv}.
    
    
    We performed the same analysis for the number of heat wave days per year as measured by the threshold-based definition. The 10 and 50-year return time maps for this definition are shown in
    \cref{fig:RT=10yrs_anomaly_RR,fig:RT=50yrs_anomaly_RR} in \cref{sec:supmat_figures}.
    The regional response patterns are similar to those for the intensity of 14-day heat waves, leading to the same conclusions.

\section{Summary of results}
\label{sec:SummaryOfResults}

    In this study, we use numerical experiment outputs from the Decadal Climate Prediction Project
    to assess which slow driver, between the AMV and spring soil moisture in Southern Europe, has the most influence on European summer heat waves. 
    We study the influence of a one-sigma deviation from the climatology for both drivers.
    We use two complementary definitions of heat waves and study both typical heat waves with return times of one or two years, and rarer events with return times of 10 or 50 years.

    In order to decouple the effects of the two drivers, we first investigated the influence of the AMV on the distribution of spring soil moisture averaged over Southern Europe.
    We found it to be negligible in two models (IPSL and CNRM), while there is a small but significant influence in EC-Earth, where the average soil moisture is 10.5 kg.m$^{-2}$ lower in the AMV+ phase compared to the AMV- phase (\cref{fig:mrso_CDF}). This can be compared to the standard deviation computed over the two ensembles together, which is $\sigma^{\textrm{SM}}_{\textrm{EC-Earth}}=34.4 \textrm{ kg.m}^{-2}$. 
    We conclude that the two drivers can be studied independently in IPSL and CNRM. In EC-Earth, there is an indirect influence of the AMV on heat waves through decreased soil moisture in the positive AMV phase, and we paid attention to removing this influence when building our Dry and Wet ensembles.

    Our main conclusion is that spring soil moisture in Southern Europe is a slow driver of greater importance than the AMV for European summer heat waves, both in terms of extension of the region of influence and in terms of amplitude. 
    While the influence of the AMV concentrates around the Mediterranean Basin, the one of spring soil moisture extends over most of continental Europe, up to 50°N or the Baltic Sea, depending on the model. 
    The amplitude of the response goes up to 4 more heat wave days per year and +0.5°C for the mean intensity of 14-day heat waves, except for EC-Earth which presents a much larger influence of spring soil moisture than the other two models
    This can be linked to the higher persistence of this variable along the summer season, as shown in \cref{fig:correlation-mrso}, and might be due to its land surface model. However, we could not discriminate with confidence if one model was more biased than the others. Further investigations need to be done in this and other models to gain a more confident understanding of the response of heat waves to spring soil moisture anomalies.
    
    Consistently with a previous study \cite{qasmi_modulation_2021}, a positive AMV phase or low soil moisture generally induce more heat wave days and hotter heat waves. 
    Regarding the AMV, all models also show a region of negative influence around Poland with responses down to -0.2°C for the mean temperature of 14-day heat waves (\cref{fig:mean_anomaly_Max14dAv}) and 1 fewer heat wave day per year (\cref{fig:mean_anomaly_RR}).
    Although this anomaly does not pass the statistical test, its presence in all three models suggests the existence of a true signal.
    We speculate that the increased surface temperature in central Europe creates a positive geopotential height response at mid troposphere (visible in \cref{fig:all_fields_JJA_response}).  This also creates a wave-like response in geopotential, which would explain the cold area around Poland.
    However, we acknowledge that our findings might be biased due to the small number of climate models used and the fact that they show inconsistent responses of physical mechanisms to the AMV (\cref{fig:all_fields_JJA_response}).

    By introducing return time maps, we were able to study extreme heat waves with return times of 10 or 50 years.
    One main issue in the study of such rare events is the scarcity of data, which causes large areas to be below the statistical significance level in \cref{fig:RT=10yrs_anomaly_Max14dAv} and \cref{fig:RT=50yrs_anomaly_Max14dAv}.
    This calls not only for new methodologies but also for longer datasets.
    For events with return times of 10 years, the influences of the AMV and soil moisture increase, according to rather similar regional patterns as for typical events, and remain of similar amplitude.
    However, the regions where a positive AMV phase or low spring soil moisture induce fewer heat wave days and cooler heat waves extend.
    This conclusion is valid for both definitions of heat waves.
    Positive AMV phases or spring soil moisture deficit induce a 1°C increase of the temperature. They also increase the number of heat wave days by up to 9 days/year.
    The return time maps for 50-year events are much noisier due to the fact that corresponding events are in the very tail of the empirical distribution.
    Once again, the influence of spring soil moisture on these extreme events is greater in EC-Earth than in the other two models.

    \section{Discussion}\label{sec:discussion}

    Our results show that the AMV can modulate the amplitude of 14-day heat waves over Southern Europe by 0.5°C to 1°C. This can be put into perspective with the current warming rate of hot extremes in Europe. Vautard et al. \cite{vautard_heat_2023} showed that the warming rate of hot extremes (TXx) over Europe is comprised between 2°C and 6°C per global warming degree with the fastest warming being over Western Europe. The current global warming rate being about 0.2°C per decade, this corresponds to a warming of 0.4°C to 1.2°C per decades for hot extremes, which is of the same order of magnitude as the modulation by the AMV. 
    This means that the phase of the AMV can either mask or exacerbate the warming trend of hot extremes.

    A similar conclusion is discussed in \cite{line_cassou} for the seasonally averaged temperature: the internal variability, including the AMV, will modulate the global warming trend at the European scale, either mitigating or exacerbating its effects. Moreover, these authors argued that, in the near term, the internal variability of the climate system will have a stronger influence on the European averaged temperature than the SSP emission scenario that we will actually follow.
    It is a natural question to ask whether this might be true also for the case of extreme heat waves, as studied in this paper. However, this is beyond the scope of our article, as it would require a specific study of the impact of the different scenarios on extreme event statistics. We leave this question as an interesting perspective for future works.
    It has to be mentioned that no distinction is made in this study between seasonal and multidecadal timescale SSTs anomalies which present different patterns. This could also be addressed in future works.

    In a warming climate, the Mediterranean Basin is projected to become drier \cite{masson-delmotte2023}. This suggests that, when considering the effect of climate change, the variability of soil moisture over the zone considered for this study might be reduced, leading to a reduced importance in the modulation of heat waves. 
    However, the transition zone between dry and wet climates, in which the variability of soil moisture is large, is expected to shift northward with respect to the Mediterranean Basin. 
    Thus, it could be relevant to look for the region of largest soil moisture variability in the future and to consider the soil moisture in this region as the relevant slow driver for European heat waves. Such a prospective study would be a natural follow-up to this work.

    As already mentioned above, a limitation of this study arises from the scarcity of data that we face when we want to study really rare events. We are already encountering this data scarcity issue for events with return times of a few decades, and the problem would be even more challenging if one wishes to study events with higher return times, for instance, a century. Recently, rare event algorithms have been designed to enhance the sampling of extreme events in numerical simulations at low computation cost. This class of algorithms has been successfully applied to sample heat waves with return times of a century or even tens of millennia in some regions of Europe \cite{ragone_computation_2018, ragone_rare_2021} and South Asia \cite{lepriol_using_2024}. Using such rare event simulations could be a promising path to study the influence of slow drivers on heat waves with return times of a century or more.
    
    The approach presented in this paper is not limited to the study of slow drivers for extreme heat waves in Europe. It could be useful for studying other rare phenomena with high impacts, and in other world regions.

\section*{Acknowledgements}
The authors thank the modelling groups who took part in the DCPP-C AMV experiment. 
The model data were accessed through the IPSL mesocenter ESPRI facility which is supported by CNRS, UPMC, Labex L-IPSL, CNES and Ecole Polytechnique.
We also thank the Centre Blaise Pascal of the Ecole Normale Superieure de Lyon for the computation resources used to carry on the study and Emmanuel Quemener for his help with the platform.
We are grateful to C. Ardilouze for suggesting to plot the soil moisture auto-correlation and bringing GLEAM dataset to our attention. We thank C. Cassou, S. Qasmi and H. Douville for enlightening discussions.

\section*{Conflict of Interest}
The authors declare no conflicts of interests.

\section*{Study fundings}
V.M. is part of the EDIPI ITN, which has received funding from the European Union’s Horizon 2020 research and innovation program under the Marie Skłodowska-Curie grant agreement No 956396.
C.L.P. is funded by the ANR grant SAMPRACE, Project No. ANR-20-CE01-0008-01.

\section*{Author Contributions}
F.B. and F.D.A. proposed the project and provided guidance for the research. 
V.M. and C.L.P. conducted the research, performed all analysis and wrote the first draft of the manuscript.
All authors contributed to manuscript review and editing.

\section*{Data Availability}
All data used in this article are publicly available.
Model simulation results can be accessed via the ESGF grid, for instance at \href{https://esgf-node.ipsl.upmc.fr/search/cmip6-ipsl/}{https://esgf-node.ipsl.upmc.fr/search/cmip6-ipsl/}. 
ERA5-Land monthly dataset, was downloaded from Copernicus Climate Data Store:
\href{https://cds.climate.copernicus.eu}{https://cds.climate.copernicus.eu}.
The GLEAM evaporation dataset can be downloaded from \href{https://www.gleam.eu/}{https://www.gleam.eu/}.
The processed data used in this research will be shared on reasonable request to the corresponding author.

\bibliography{Bibliography/AMV-heatwaves}

\newpage

\appendix

\section{Supplementary material}
\label{sec:supmat}
    \subsection{Methods}\label{sec:supmat_method}

        \paragraph{Brief descriptions of the models characteristics} 
        \begin{itemize}
            \item \textit{IPSL} is composed of the LMDZ atmospheric model version 6A-LR, based on a rectangular grid with 144 longitude x 143 latitude equally spaced points, resulting in a resolution of 2.5° × 1.3°. It has 79 vertical levels and extends up to 80 km. The ocean component is the NEMO oceanic model version 3.6, with 71 vertical layers and an horizontal resolution of 1°. The land surface model is the ORCHIDEE version 2.0, with 11 layers for a total of 2 m of soil depth. Further information and details can be found at \cite{boucher_presentation_2020}.  
            \item \textit{EC-Earth} is composed of the Integrated Forecast System (IFS) CY36R4 of the European Centre for Medium Range Weather Forecasts (ECMWF) atmospheric model, based on a linearly reduced Gaussian grid equivalent to 512 longitude x 256 latitude points with 91 vertical levels. It includes the land-surface scheme HTESSEL. The ocean and sea-ice model is NEMO-LIM3 version 3.6, with 75 vertical layers.  Further information and details can be found at \cite{doscher_ec-earth3_2021}.
            \item \textit{CNRM} is composed of the ARPEGE-Climat atmospheric model, version 6.3 with 91 vertical levels. The ocean component is based on the NEMO version 3.6, while the sea ice component is based on GELATO, version 6. Further information and details can be found at \cite{voldoire_evaluation_2019}.
        \end{itemize}
        
\begin{figure}
        \centering
        \includegraphics[width=0.6\textwidth]{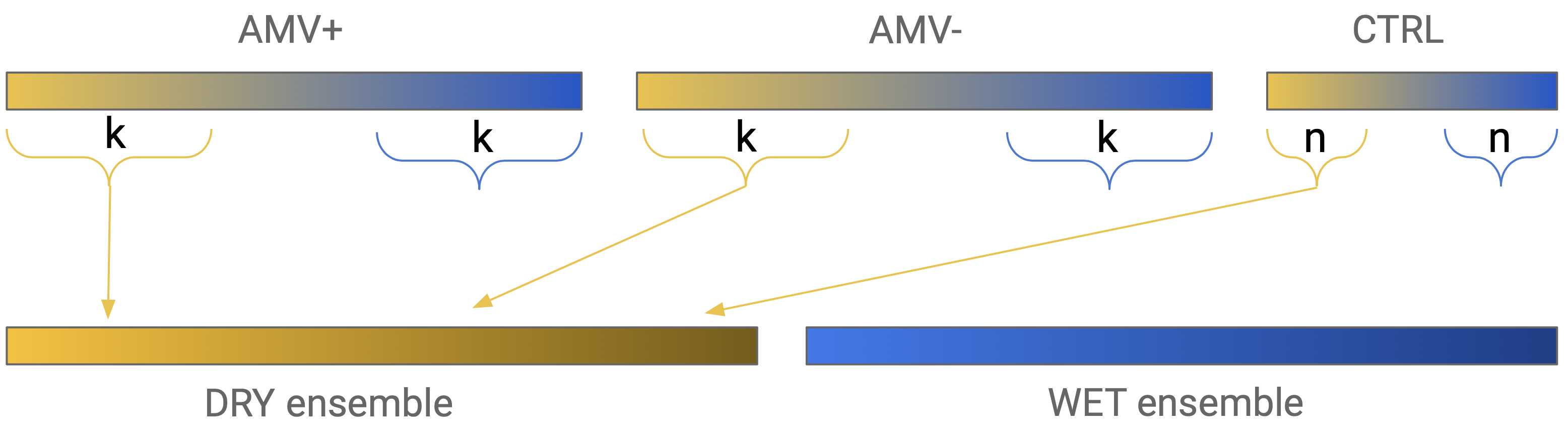}
        \caption{\label{fig:Dry/wet_construction} \textit{Sketch of the procedure to create the Dry and Wet ensembles.} Data in each experiments are already sorted according to its $\text{SM}_{\text{av}}$ value.}
        \end{figure}

        \paragraph{Creation of the Dry and Wet ensembles}
        The Dry and Wet ensembles are made of years coming from the AMV+/- and CTRL ensembles according to $\text{SM}_{\text{av}}$. When building the Dry and Wet ensembles we want to make sure that there is no indirect influence of the AMV on the soil moisture through influence of the AMV on $\text{SM}_{\text{av}}$. \Cref{fig:mrso_CDF} shows the distribution of $\text{SM}_{\text{av}}$ for each ensemble in the three models. We see that in IPSL and CNRM the phase of the AM does not influence the soil moisture, while in EC-Earth, there is a non negligible influence with the average $\text{SM}_{\text{av}}$ in the AMV- phase being 5.2 $kg.m^{-2}$ than in the AMV+ phase. 
        To make sure that there is no indirect influence of the AMV we choose to enforce to have the exact same numbers of AMV+ and AMV- years in the Dry and Wet ensembles.This choice also allows us to deal with the imbalance of the CNRM ensemble because of missing runs for Soil Moisture data. We have only 380 years of AMV+ and 280 years of AMV-. We will briefly comment on the procedure to create the ensemble. It is sketched in \cref{fig:Dry/wet_construction}. Each original ensemble (AMV+, AMV-, CTRL) is sorted according to $\text{SM}_{\text{av}}$. The k driest years of the AMV+ and AMV- ensembles are put in the Dry ensemble which is completed with the n driest years from the CTRL run when available. For this study, the number of years coming from the AMV experiments, namely k, is different from the one coming from the CTRL, namely n. The reason is the different amount of years available for each experiment, which we detail in \cref{tab: Ensemble_sizes}. 
        More details about the values of $\text{SM}_{\text{av}}$ for each model can be found in \cref{tab: SM_av in Dry/Wet realized vs targer}.
        We choose the ensemble size such that the mean value of $\text{SM}_{\text{av}}$ in the Dry (resp. Wet) ensemble is nearly one standard deviation below (resp. above) the mean value $\text{SM}_{\text{av}}$ averaged over the AMV+/- and CTRL ensembles all-together. 
        This makes the comparison with the AMV forcing sensible because the imposed SST pattern corresponds to one standard deviation of the AMV variability. 
         \begin{center}
        \begin{table*}
        \centering
        	\begin{tabular}{|c|c|c|c|}
        		\hline
        		    & IPSL         & EC-Earth     & CNRM  \\ \hline
        		Dry & 443.1 / 442.5 & 715.2 / 714.3 & 383.1 / 383.2  \\ \hline
        		Wet & 480.2 / 481.0 & 782.5 / 783.1 & 404.6 / 405.7  \\
        		\hline
        	\end{tabular}
        	\caption{\textit{Mean value of $\text{SM}_{\text{av}}$ (in $kg.m^{-2}$) in the Dry and Wet ensembles for each model.} 
        	In each case the values are ordered as \textit{realized / target} where \textit{realized} is the computed mean in the Dry or Wet ensemble and  \textit{target} corresponds to exactly one standard deviation away from the overall mean value. 
        	 \label{tab: SM_av in Dry/Wet realized vs targer}}
        \end{table*}
        \end{center}
        
    \paragraph{Computation of the errorbars on the local RT plots}

    Let $A$ be the random variable of event's amplitudes. Let us fix a return level $a$.
    The random variable $X = \mathbf{1}(A \geq a)$ follows a Bernoulli law of parameter
    $p:= \mathbb{P}(A \geq a)$
    Let $N$ be the number of years in our dataset and $m$ the number of years where we observe $A\geq a$. Then $\hat{p}= m/N$ is an unbiased estimator of $p$ and we estimate the return time of events larger than $a$ as 
    \begin{equation}\label{eq: return time formula}
        \hat{r}(a) = \frac{1}{\hat{p}} =\frac{N}{m}.
    \end{equation}

\begin{figure}
    \centering
    \includegraphics[width=1\textwidth]{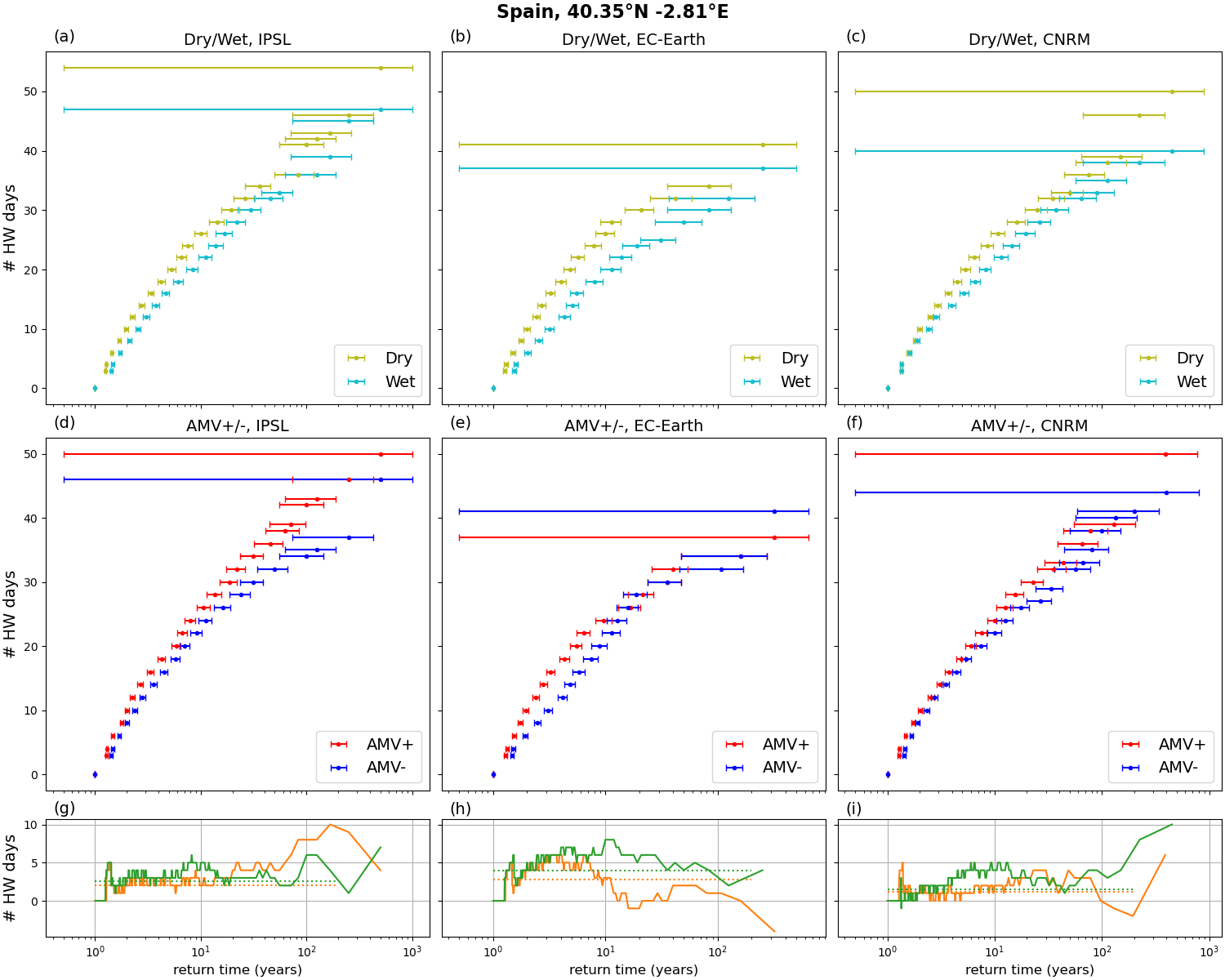}
    \caption{\label{fig:RT_Spain_RR} \textit{Return time curves for the threshold based definition (\cref{def:hw-threshold})}. The plots show the yearly number of heatwave days as a function of the return time (in years) for Dry and Wet ensembles (top row) and for AMV+ and AMV- ensembles (bottom row) for the three models. Error bars correspond to one standard deviation of the estimated return time needed to observe an event of a certain amplitude.  The third row shows the Dry - Wet (green) and AMV+ - AMV- (orange) differences as a function of the return time. The dotted lines indicate the mean difference values. For each model the chosen grid point covers the coordinate indicated in the figure title which corresponds to a location in the Northeast of Spain, close to the Mediterranean Sea.}
    \end{figure}

    We want to compute an estimate of the error on $\hat{r}(a)$.
    First, note that the empirical variance of $\hat{p}$ reads 
    $\hat{\sigma}_X^2 = \frac{1}{N-1}\sum_{i=1}^N(X_i-\hat{p})^2 = \frac{N}{N-1}\hat{p}(1-\hat{p})$.
    According to the Central Limit Theorem, the variance of $\hat{p}$ reads
    $\sigma_{\hat{p}}^2 = \frac{p(1-p)}{N}$ which can be estimated by the empirical formula
    $$\hat{\sigma}_{\hat{p}}^2 = \frac{\hat{p}(1-\hat{p})}{N} \, .$$
    Let us define $\delta = \hat{p} - p$.
    Note that $\mathbb{E}[\delta]=0$ and $\sigma^2_{\delta} = \sigma_{\hat{p}}^2$
    Let us assume that in all our observations $\delta < p$ (which is likely to be valid if $\sigma(r) \ll r$). Then we can write:
    \begin{align}
    \hat{r} &= \frac{1}{\hat{p}} = \frac{1}{p} \times \frac{1}{1 + \delta/p} = 
    		\frac{1}{p} \left(1 - \frac{\delta}{p} + \frac{\delta^2}{p^2} + O\left( \frac{\delta^3}{p^3} \right) \right) \, , \\
    \hat{r}^{2} &= \frac{1}{n^{2}_+} = \frac{1}{p^2} \left(1 - \frac{2\delta}{p} + \frac{3\delta^2}{p^2} + O\left( \frac{\delta^3}{p^3} \right) \right) \, .
    \end{align}
    Taking the difference of the expectations we get:
    \begin{align}
    \sigma^2_{\hat{r}} &=  \mathbb{E}[\hat{r}^2] - \mathbb{E}[\hat{r}]^{2} = \frac{\mathbb{E}[\delta^2]}{p^4} + O\left( \frac{\mathbb{E}[\delta^3]}{p^3} \right) \, . 
    \end{align}
    The errorbars plotted on the local return time plots \cref{fig:RT_Spain_Max14dAv,fig:RT_Spain_RR} corresponds to the standard deviation of $\hat{r}$:
    \begin{equation}
    \sigma_{\hat{r}} \simeq \frac{\sigma(\hat{p})}{p^2} = \frac{1}{p^2} \sqrt{\frac{p(1-p)}{N}} \simeq \hat{r} \sqrt{\frac{\hat{r}-1}{N}} \, ,
    \end{equation}
    where we have used that $1/p=r$.

    Note that the domain of validity of the this approximation is $\sigma(r) \ll r$ and so these formula is not correct to estimate the error on the most extreme return levels (the ones for which 
    the number of exceedance $m$ is small).

    \paragraph{Implementation of the bootstrap test}
    We implemented a bootstrap to test the significance of the AMV+ - AMV- and Dry-Wet differences in all maps. The procedure is the same for each model, but was slightly adapted for CNRM to take into account the fact that the AMV+ and AMV- do not have the same size and that the soil moisture of some members is missing. 

    Concretely, we pooled the results from all experiments (AMV+/- and CTRL) together to obtain a single large dataset for each model. From this large dataset, we draw N=1000 samples of M years, where M is equal to the number of year in each dataset displayed in \cref{tab: Ensemble_sizes}.
    For each sample we compute the average over the sample, as well as the 10-year and 50-year return time maps. 
    For each of those maps (average, 10-year and 50-year return time maps), we build the empirical distribution of all differences between two distinct samples and compute the quantiles from the distribution.
    We consider the difference at each grid point to be significant at the 95\% significance level whenever it is lower than the quantile 0.025 or higher than the quantile 0.975.

    \subsection{Additional figures}\label{sec:supmat_figures}

    \begin{figure}
    \centering
    \includegraphics[width=1.15\textwidth]{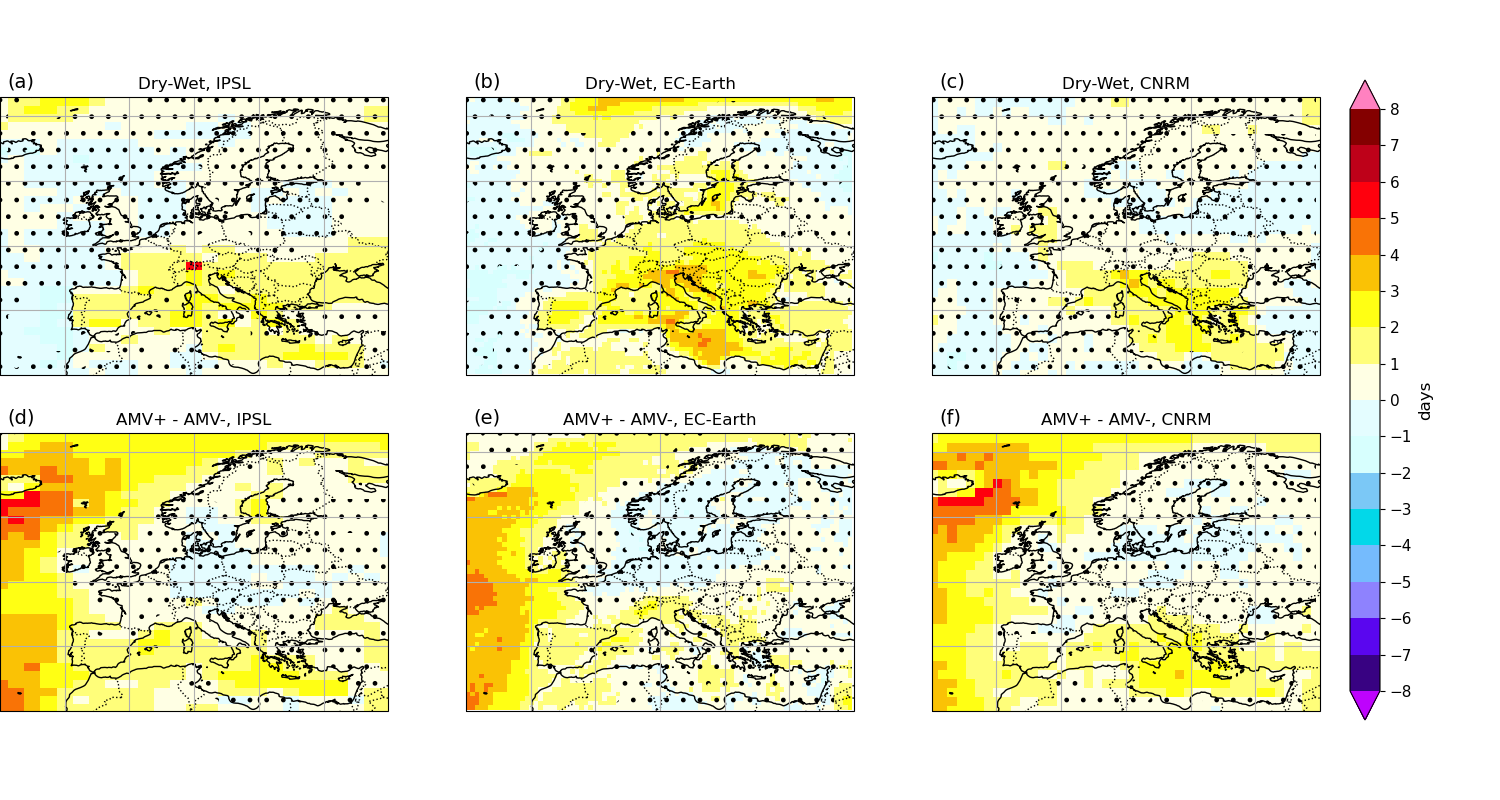}
    \caption{\label{fig:mean_anomaly_RR_9580} \textit{Mean anomaly maps for the threshold based definition}, with thresholds $T_1$ and $T_2$ corresponding to the 95th and 80th percentile of the local JJA $T_{max}$ distribution. The maps show the difference of the mean number of heatwaves days per year between (top line) the dry and the wet ensembles and (bottom line) the AMV+ and AMV- ensembles for each of the three models. Hatching denotes area below the 95\% significance level according to a bootstrap test.}
    \end{figure}

    \begin{figure}
    \centering
    \includegraphics[width=1\textwidth]{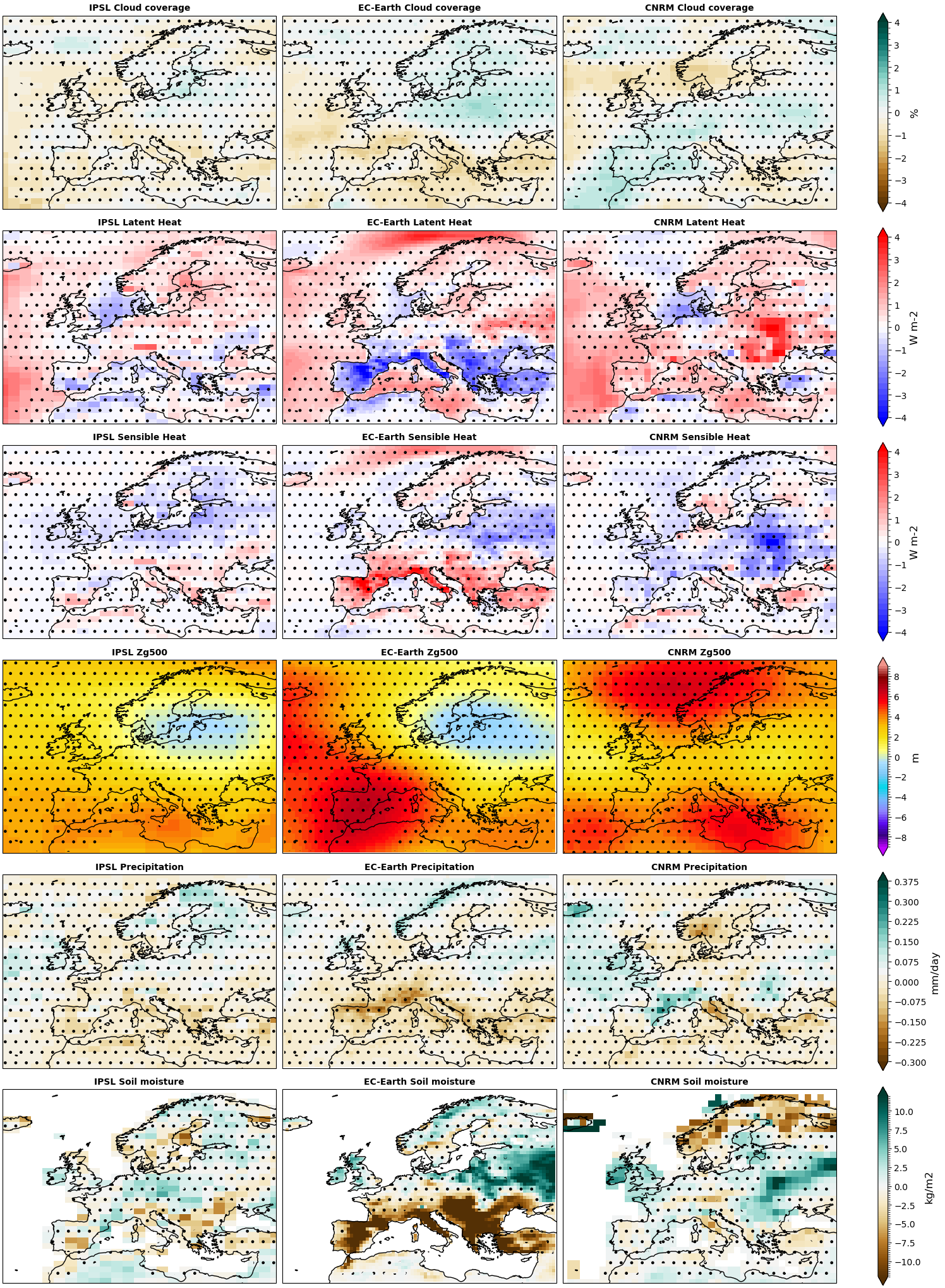}
        \caption{\label{fig:all_fields_JJA_response} \textit{Mean response maps to the AMV phase for different fields.} Each map represents the mean difference between AMV+ and AMV- of (from top to bottom row) percentage of cloud coverage, Latent heat, sensible heat, 500hPa geopotential height, precipitation, soil moisture. The average is taken with respect to the months of June, July and August. Each column represents a model. Stippling denotes area below the 95\% significance level according to a bootstrap test.}
    \end{figure}

    \begin{figure}
    \centering
    \includegraphics[width=1\textwidth]{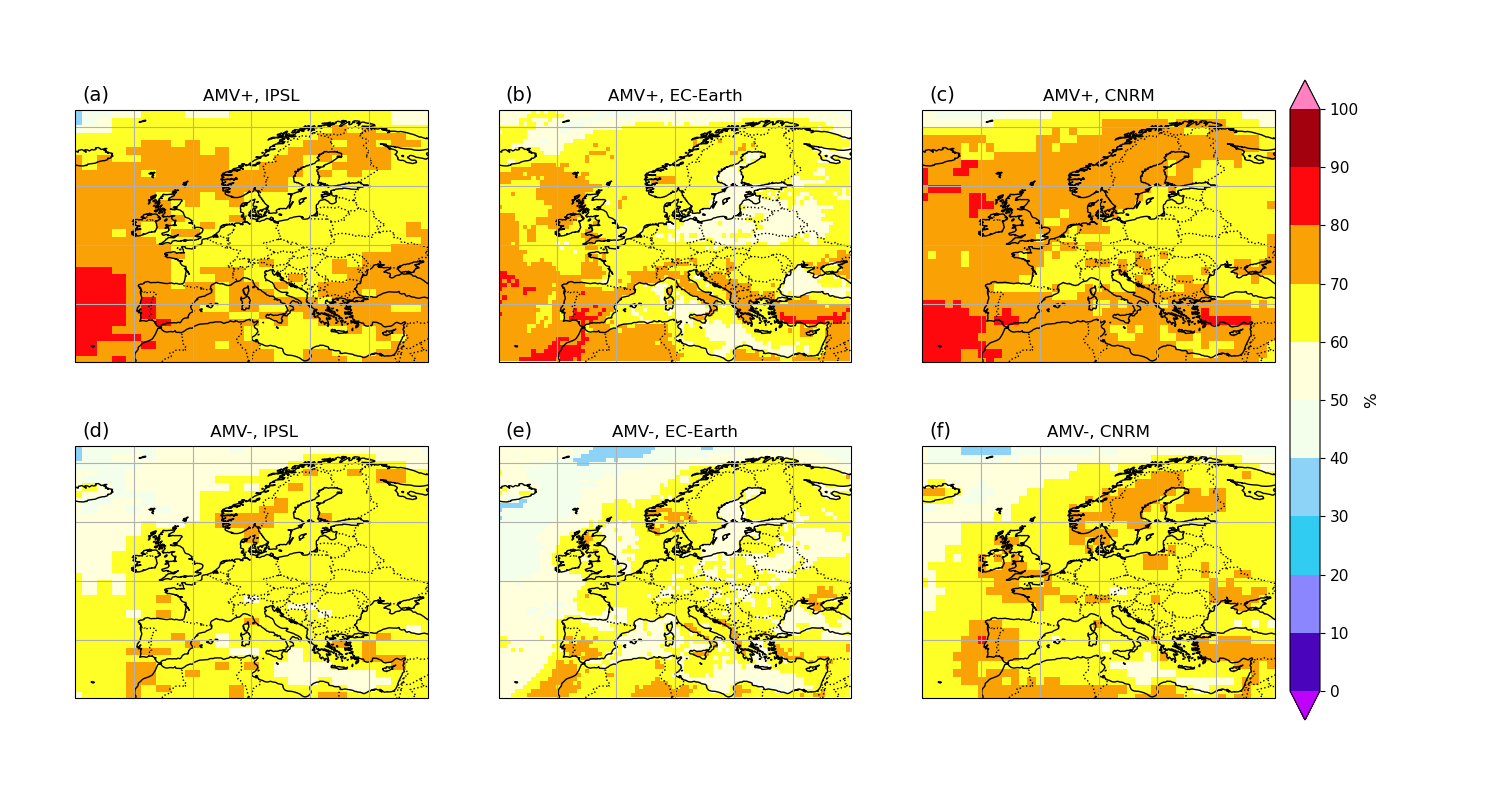}
        \caption{\label{fig:frac_HW_yearT9075} \textit{Percentage of years during which at least one threshold-based heat wave occurs.}
        These maps show the percentage of years during which at least one heat wave (according to the threshold-base definition, see \cref{def:hw-threshold}) occurs. This percentage is above 50\% over all land area, in all model and in the two AMV phases. Since these heat waves occur more frequently than every two years, they have a return time of less than two years. The percentage is even larger than 60\% over most of Europe and 70\% in many locations.}
    \end{figure}
    
    \begin{figure}
    \centering
    \includegraphics[width=1.05\textwidth]{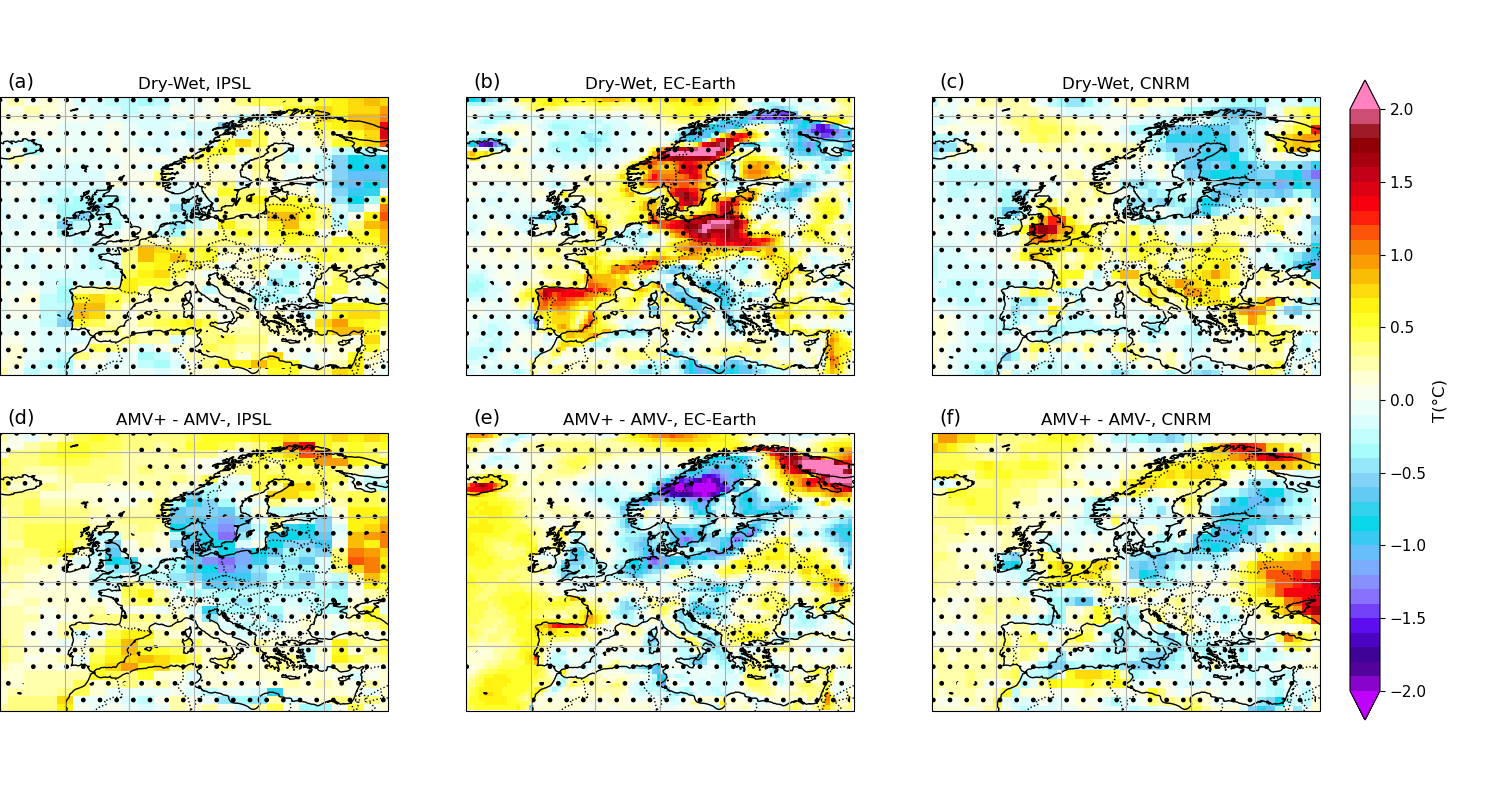}
    \caption{\label{fig:RT=50yrs_anomaly_Max14dAv} \textit{50 years return time maps for 14-day heat waves.} The maps show the  difference of $\tilde{A}$ between (top line) the Dry and Wet ensembles and (bottom line) the AMV+ and AMV- ensembles, conditioned over events with return time greater than 50 years. Stippling denotes area below the 95\% significance level according to a bootstrap test.}
    \end{figure}

    \begin{figure}
    \centering
    \includegraphics[width=1.15\textwidth]{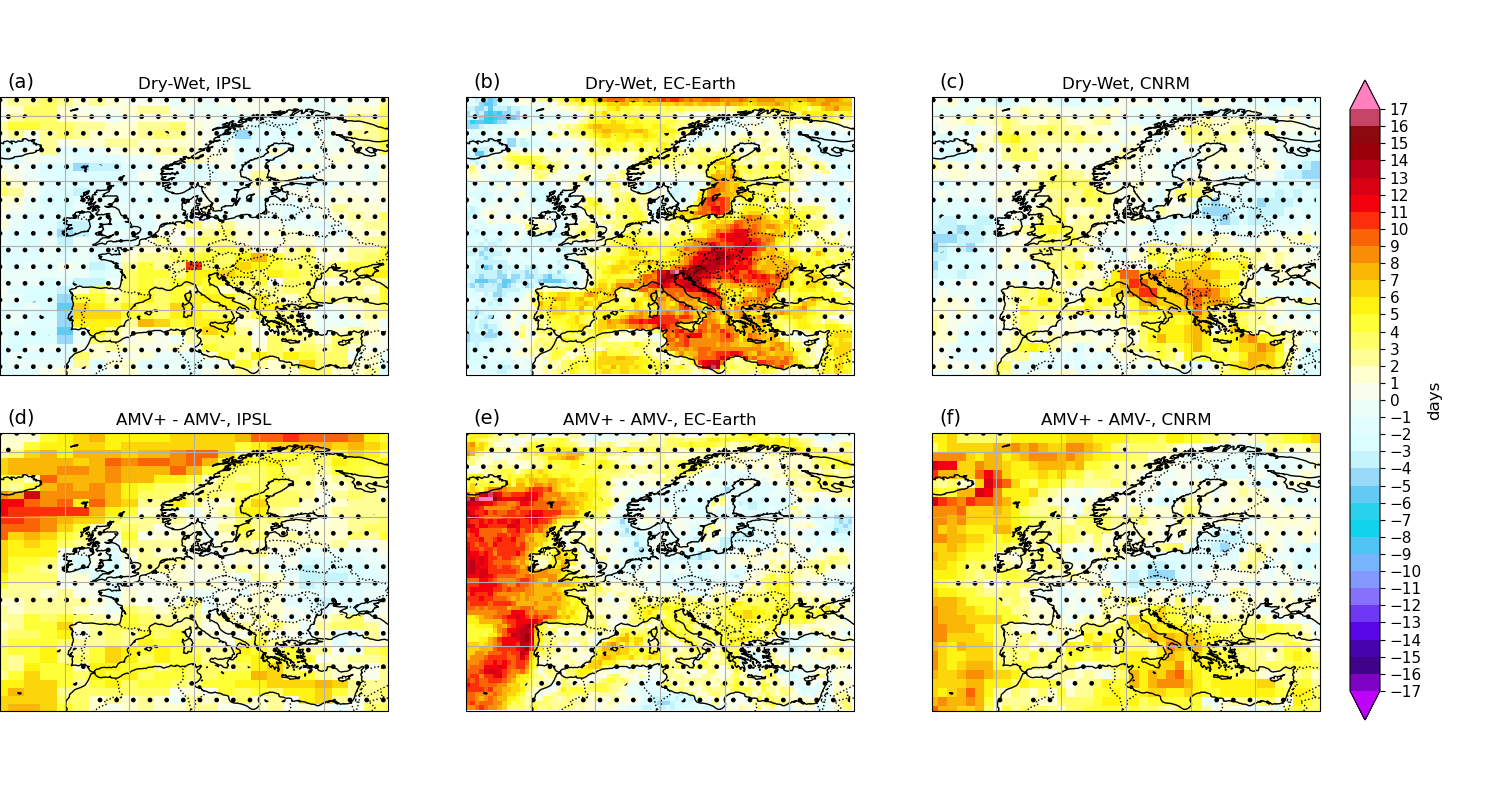}
    \caption{\label{fig:RT=10yrs_anomaly_RR} \textit{10 years return time maps for the threshold based definition}. The maps show the yearly heatwaves days difference between (top line) the dry and wet ensembles and (bottom line) the AMV+ and AMV- ensembles, conditioned over the return time of 10 years. Stippling denotes area below the 95\% significance level according to a bootstrap test.}
    \end{figure}

    \begin{figure}
    \centering
    \includegraphics[width=1.15\textwidth]{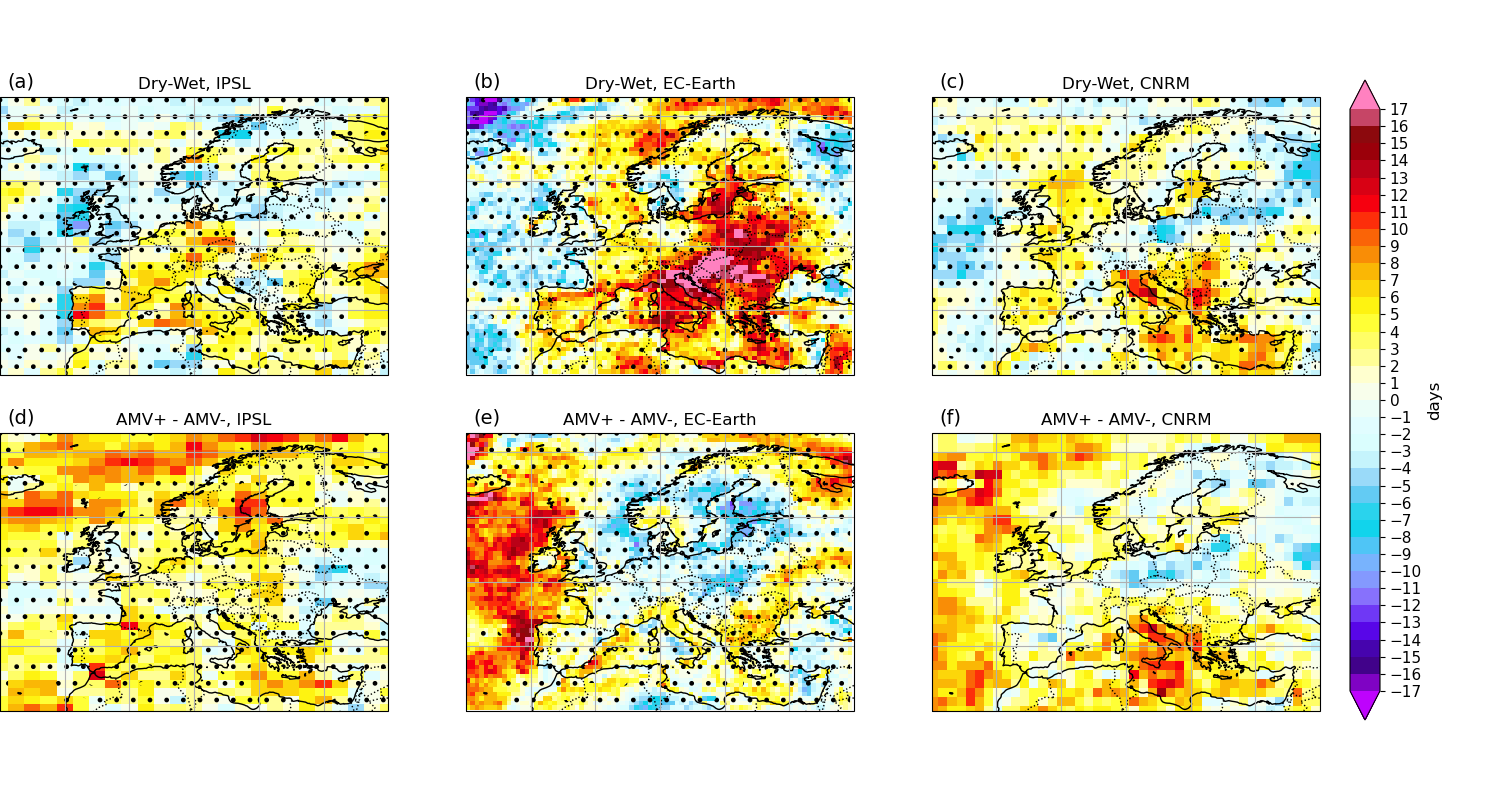}
    \caption{\label{fig:RT=50yrs_anomaly_RR} \textit{50 years return time maps for the threshold based definition}. The maps show the yearly heatwaves days difference between (top line) the dry and wet ensembles and (bottom line) the AMV+ and AMV- ensembles, conditioned over the return time of 50 years. Stippling denotes area below the 95\% significance level according to a bootstrap test.}
    \end{figure}

\end{document}